\documentclass[11pt]{amsart}
\usepackage{caption,subcaption}

\usepackage{graphicx}
\usepackage{amssymb}
\usepackage{epstopdf}

\usepackage{placeins}
\usepackage{secdot}
\usepackage{multirow}
\usepackage{placeins}
\usepackage{soul}
\usepackage{rotating}
\usepackage{epsfig}

\usepackage{pgf}

\usepackage{url}
\usepackage{pgfpages}

\usepackage{geometry}                
 \usepackage{array}
\usepackage{subfig}
\usepackage[compact]{titlesec}

\usepackage{enumitem}      
   \newenvironment{compenum}{\begin{enumerate}[topsep=0pt,itemsep=0pt,parsep=0pt,partopsep=0pt]}{\end{enumerate}}

\newtheorem{lemma}{Lemma}[section]
\newtheorem{cor}{Corollary}[section]
\newtheorem{claim}{Claim}[section]

\newtheorem{thm}{Theorem}[section]
\newtheorem{theorem}{Theorem}[section]
\newtheorem{remark}{Remark}[section]
\newtheorem{comment}{Comment}[section]

\newcommand{\taba}{\hspace*{0.25in}}
\newcommand{\tabb}{\hspace*{.5in}}

\def\half{\mbox{$\frac{1}{2}$}}
\def\23{\mbox{$\frac{2}{3}$}}

\def\x{\mathbf{x}}

\def\d{\mathbf{d}}
\def\f{\mathbf{f}}

\def\0{\mathbf{0}}
\def\E{\mathbf{ET}}
\def\L{\mathbf{LT}}

\begin{document}

\title[An Efficient All-Min-Cuts algorithm]{
Min cost flow on unit capacity networks and convex cost K-flow are as easy as the assignment problem with All-Min-Cuts algorithm}


 \author{Dorit S. Hochbaum
 }
\thanks{Department of Industrial Engineering and Operations Research,
University of California, Berkeley, CA 94720
(dhochbaum@berkeley.edu). This author's research was supported
in part by NSF award No. CMMI-1200592.
}

\begin{abstract}
We explore here surprising links between the time-cost-tradeoff problem and the minimum cost flow problem that lead to fast, strongly polynomial, algorithms for both problems.  One of the main results is a new algorithm for the unit capacity min cost flow that  matches the complexity of the fastest strongly polynomial algorithm known for the assignment problem.

The time cost tradeoff (TCT) problem in project management  is to expedite the durations of activities, subject to precedence constraints, in order to achieve a target project completion time at minimum expediting costs, or, to maximize the net benefit from a reward associated with project completion time reduction. Each activity is associated with integer normal duration, minimum duration, and expediting cost per unit reduction in duration.  We devise here the {\em all-min-cuts} procedure,  which for a given maximum flow, is capable of generating all minimum cuts (equivalent to minimum cost expediting) of equal value very efficiently.  Equivalently, the procedure identifies all solutions that reside on the TCT curve between consecutive breakpoints in average $O(m+n \log n)$ time, where $m$ and $n$ are the numbers of arcs and nodes in the network.

The all-min-cuts procedure implies faster algorithms for TCT problems that have ``small" number of breakpoints in the respective TCT curve: For a project network on $n$ nodes and $m$ arcs, with $n'$ arcs of finite uniform expediting costs, the run time is $O((n+n')(m+n\log n))$; for projects with {\em rewards} of $O(K)$ per unit reduction in the project completion time the run time is $O((n+K)(m+n\log n))$.

Using the primal-dual relationship between TCT and the minimum cost flow problem (MCF) we generate fast strongly polynomial algorithms for various cases of minimum cost flow:  For MCF on unit (vertex) capacity network, the complexity of our algorithm is $O(n(m+n\log n))$;
For a minimum convex (or linear) cost $K$-flow problem our algorithm runs in $O((n+K)(m+n\log n))$; for MCF on $n'$ constant capacities arcs, the all-min-cuts algorithm runs in $O((n+n')(m+n\log n))$ steps.
This complexity of the algorithm for any min cost $O(n)$-flow matches the best known complexity for the {\em assignment problem},  $O(n(m+n\log n))$, yet with a significantly different approach.  
{\bf In this revision, version 2, we note that the known successive shortest paths algorithm matches the complexity of the all-min-cuts algorithm albeit with a very different approach. 
In version 1 of this paper, the author failed to recognize that the successive shortest paths algorithm runs as fast as the all-min-cuts algorithm for the type of min cost flow problems discussed here.}

\end{abstract}

\maketitle

\section{Introduction}

We present here new algorithms for the time cost tradeoff problem in project management, and explore surprising links between the time-cost-tradeoff problem in project management and the minimum cost network flow problem.  The two problems are linear programming duals of each other as has been known for a long time: In 1961 Fulkerson and Kelley, independently, \cite{Ful61,Kel61}, recognized that the dual of the linear programming formulation of the time-cost-tradeoff (TCT) problem is a form of a min cost flow  (MCF) problem.  This they used to establish that the TCT problem can be solved via the dual using flow techniques with the out-of-kilter method, known at the time, which does not run in polynomial time.
The surprise here is that the new algorithms we devise for the TCT problem are not only faster for various cases of TCT, but also imply considerable improvements in the complexity of algorithms for well known MCF problems, including the minimum cost unit capacity network flow and the minimum cost $K$-flow for which the amount of flow through the network is $K$.   Furthermore, our new methodology departs significantly from any methods currently used for MCF problems, and as such it affords fresh insights in this well studied area, with a potential for additional improvements in complexities beyond the ones reported here.  We refer to our algorithm as the AMC-algorithm after the key procedure of {\em all-min-cuts} (AMC).  The complexity of our algorithm depends on the number of breakpoints in the time cost tradeoff curve (TCT curve) in the dual which, for a $K$-flow problem, cannot exceed $K$.

The time cost tradeoff (TCT) problem in project management is to expedite the durations of activities, subject to precedence constraints, in order to achieve a target project finish time at minimum expediting costs, or, to maximize the net benefit from a reward associated with project completion time reduction.    Each activity is associated with normal duration, minimum duration, and expediting cost per unit reduction in duration.  An important concept is that of the  {\em TCT curve}.  This is a function that maps each project finish time to the minimum expediting costs required to attain that finish time.  The key new idea here is a new algorithm, the {\em all-min-cuts} algorithm,  which, for a given vector of durations and a given maximum flow on a corresponding network, it finds all the minimum cuts of value equal to that of the maximum flow, each of which corresponds to minimum cost reduction in the project finish time.   For a project network $G=(V,A)$ with $n=|V|$ and $m=|A|$, the algorithm generates the minimum cost expediting solution for every finish time value until the subsequent breakpoint in the {\em TCT curve} is reached and the cost of per-unit further reduction in finish time is strictly higher, in $O(m+n\log n)$ time.  For networks with small number of breakpoints, this run time is particularly efficient.  Such networks include project networks with uniform expediting costs, or the TCT problem of maximizing the net benefits from a reward of $K$ per unit reduction in the project finish time.  For the uniform costs project network with $n'$ activity arcs (and $m-n'$ precedence arcs), there are at most $O(n')$ breakpoints. For the $K$ reward project network there are at most $O(K)$ breakpoints.

A project network is a DAG (Directed Acyclic Graph) $G=(V,A)$ with start and finish nodes $s$ and $f$. In the AOA (Activity On Arc) representation of project network the arcs of the network are associated with activities or precedence relations.  Each activity $(i,j)\in A$ has normal duration $d_{ij}$, minimum duration $\underline{d}_{ij}$ and expediting cost $w_{ij}$. A precedence arc has $0$ duration and expediting cost $\infty$.  The ``normal" finish time, or makespan, of the project $T^0$ is the length of the {\em longest} path from $s$ to $f$.  This longest path is easily computed using a dynamic programming procedure known as CPM (Critical Path Method described in Section \ref{sec:prelim}).  One form of the time cost trade-off problem is to reduce the project finish time to $T^*<T^0$ by expediting the durations of activities so that total cost of expediting is minimum.  Other forms have budget for expediting costs, or a reward per unit reduction in the makespan of the project.  Using the notation $x_{ij}$ for the expedited duration of activity $(i,j)$, and $t_i$ the start time of event $i$, associated with node $i$, the TCT problem with target makespan $T$ is formulated as the linear program, in which the given objective function is equivalent to the intuitive objective: $\min  \sum _{(i,j)\in A} w_{ij}(d_{ij}- x_{ij})$:

\[
({\mbox{\sf LP-TCT}}) \hspace{.1in}\begin{array}{ll}
\mbox{} Z(T)=\max \ &  \sum _{(i,j)\in A} w_{ij} x_{ij} \\
\mbox{subject to }\ &  t_j-t_i \geq x_{ij} \ \ \mbox {for all} \quad (i,j)  \in A\\
& t_s=0\\
& t_f =T \\
&\underline{d}_{ij} \leq x_{ij} \leq d_{ij} \ \mbox {for all} \quad (i,j)  \in A.
\end{array}
\]

\begin{figure}[h!]
\vspace{-0.7in}
  \begin{center}
 \scalebox{0.57}
  { \hspace{0.7in}
      \epsfig{figure =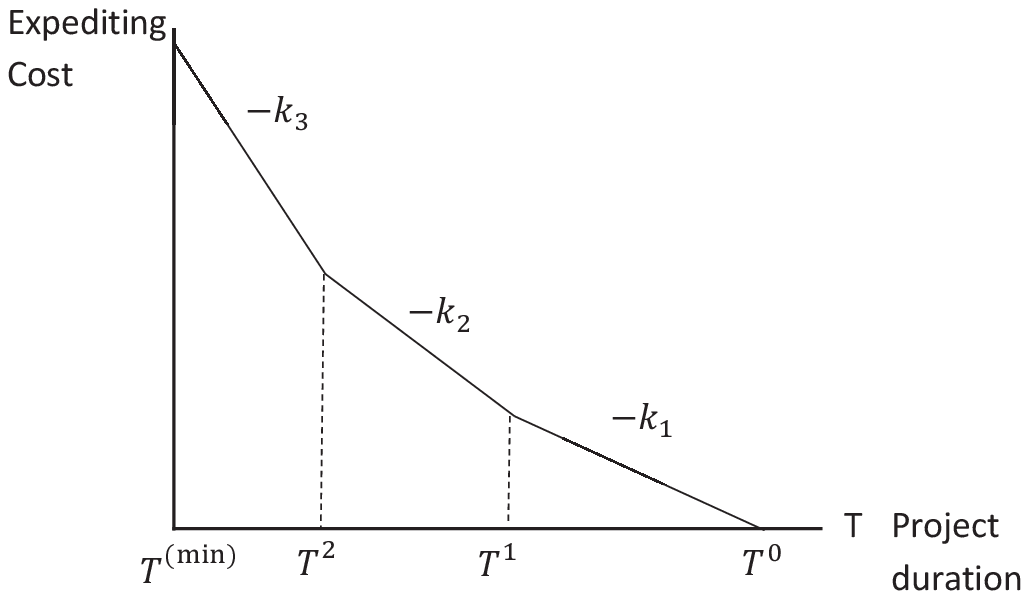}
      }
      \end{center}
\vspace{-3.21in}
  \caption{The time-cost trade-off curve.  }\label{fig:breakpoints}
\end{figure}

The {\em TCT curve} associated with a particular project network is the function $\sum_{(i,j)\in A} w_{ij} d_{ij}- Z(T)$ of the minimum expediting cost as a function of the target project duration, which is a piecewise linear convex function.  The example TCT curve function, $\sum_{(i,j)\in A} w_{ij} d_{ij}- Z(T)$, illustrated in Figure \ref{fig:breakpoints}, has 3 linear pieces and 2 breakpoints.

The TCT problem is known to be solved as a linear programming problem, and in integers, as the constraint matrix of (LP-TCT) is totally unimodular.  Other than linear programming, algorithms for the TCT problem were considered decades ago.  In 1961 Fulkerson and  Kelley, independently, \cite{Ful61,Kel61}, recognized that the dual of LP-TCT is a form of a min cost flow  (MCF) problem, and thus can be solved via the dual. Consequently, TCT is solved with any efficient algorithm available to date for MCF, e.g.\ \cite{Orl93}, and then constructing the respective dual solution by finding the single source shortest paths labels in the residual graphs.  Another algorithm to solve TCT was proposed by Phillips and Dessouky \cite{PD77}, PD-algorithm.  The PD-algorithm does not run in polynomial time, yet its technique, referred to here as {\em repeated cuts}, is fundamental to the basic structure of our AMC algorithm.

The algorithm of Phillips and Dessouky for TCT is a primal algorithm  that identifies for each project duration the least cost critical activities to crash (expedite), so as to reduce the makespan by {\em one} unit. The least cost expediting that reduces the makespan by one unit corresponds to a solution to a minimum cut problem on an associated graph.  We refer to any algorithm that reduces the makespan by using a sequence of minimum cuts (minimum $s,t$-cuts) in a graph, as a {\em repeated cuts} algorithm. The repeated cuts algorithm of Phillips and Dessouky constructs, for given activity durations,  a respective network of the critical activities and finds, using a cut procedure, the collection of activities whose crashing by one unit, or duration increase by one unit, will reduce the project duration by one unit at a minimum cost.  That is, the PD-algorithm generates for {\em each} integer project duration $T$, the respective value $Z(T)$ on the TCT-curve. (A detailed description of the algorithm of Phillips and Dessouky is given in Section \ref{sec:PD}.) This approach is inherently of pseudo-polynomial complexity, as the number of iterations is the gap $T^0-T^*$ between the normal and the target makespans.  
Recently, in \cite{Hoc15}, we devised a polynomial time variant of the algorithm of Phillips and Dessouky that runs in time $O(n\log{\frac{T-T^*}{n}}T(n,m))$ for $T(n,m)$ the run time of any minimum $s,t$-cut algorithm on a graph with $n$ nodes and $m$ arcs.  Our approach here employs repeated cuts with complexity that depends on the number of breakpoints in the TCT curve which, for certain TCT problems, is guaranteed to be ``small".

The convex cost version of the TCT problem was shown to be polynomial time solvable with flow techniques with the most efficient algorithm to date reported in \cite{AHO03}.  The gist of that algorithm is a reduction of the problem to a convex minimum cost flow (MCF) problem.  In that sense, again, the solution for the convex TCT problem is attained by considering the respective MCF dual.  For the convex cost $K$-reward TCT problem, for $K$ ``small", e.g.\ $O(n)$, we present here a more efficient, and primal, algorithm.

The minimum cost flow (MCF) problem is defined for a digraph $G=(V,A)$ with costs per unit flow $c_{ij}$ and capacity upper bounds $u_{ij}$ associated with each arc $(i,j)\in A$, and supply $b_v$ at each node $v\in V$.  Without loss of generality the capacity lower bounds can be set to $0$. A standard formulation of MCF is,

\vspace{-0.15in}
 \begin{eqnarray*}
({\mbox{\sf MCF }})  \ \    \mbox{min} &  \sum_{(i,j)\in A}\ c_{ij} f_{ij} & {} \\
     \mbox{subject to} & \sum_{k \in V, (i,k)\in A} f_{ik}\,-\,
                         \sum_{k \in V, (k,i)\in A} f_{ki} = b_i  &\ \forall i \in V \\
     {} & 0 \le f_{ij} \le u_{ij}, & 
                         \ \forall (i,j) \in A.
  \end{eqnarray*}

The first set of constraints is known as the flow balance
constraints. The second set of constraints is known as the capacity
constraints. Nodes with positive (negative) supply are
referred to as {\it supply} ({\it demand}) nodes and nodes with $0$ supply
are referred to as {\it transshipment} nodes.  For MCF to have a feasible solution, the total supply $B$ must equal the total demand, $B=\sum _{i\in V|b_i>0} b_i = \sum _{j\in V|b_j<0} -b_j$, and the network must have sufficient capacity to accommodate the flows from supply to demand nodes.  These conditions can be easily verified as discussed next.

Any MCF problem can be presented as a $s,t$ flow problem, where $s$ is a dummy node adjacent to all supply nodes with arcs of capacities $u_{si}=b_i$, and $t$ is adjacent to all demand nodes with arcs of capacities $u_{jt}=|b_j|$.  MCF is then the minimum cost flow of $B$ units from $s$ to $t$.  This flow of $B$ units must be the maximum possible flow in this $s,t$ network or else the problem is infeasible.    A more general variant of the problem requires to send from $s$ to $t$ a flow of $K$ units such that  $K\leq B$.  When $K$ is the maximum flow in this $s,t$ graph the problem is also known as the {\em max-flow min cost flow} problem.  For other, arbitrary values of $K$, the problem can be formulated as min cost flow by adding another node $s'$ adjacent to $s$ with arc $(s',s)$ of capacity lower and upper bounds equal to $K$.  Alternatively the problem can be presented as circulation, with an arc from $t$ to $s$, of capacity lower and upper bounds equal to $K$. (For the circulation MCF all supplies and demands are $0$, i.e.\ all nodes are transshipment nodes.)  We refer to this problem as {\em min cost $K$-flow} problem, or $K$-MCF.
The minimum {\em convex cost $K$-flow} problem is the min cost $K$-flow where the cost on each arc is a convex function of the amount of flow on that arc.


\subsection{Results here}
For the minimum cost flow problems discussed here the only known strongly polynomial algorithms are algorithms that apply for the general MCF.  The fastest strongly algorithm for MCF is Orlin's algorithm, of complexity $O(m\log n (m+n\log n))$ for a graph on $n$ nodes and $m$ finite capacity arcs, \cite{Orl93}.  We refer to this run time as $MCF(n,m)$.

For the min cost $K$-flow there are no known previous algorithms that exploit the limit on the amount of flow, or equivalently, on a bound on the capacity of a minimum cut in the graph $G=(V,A)$.  The well known and much studied assignment problem is a special case of min cost $K$-flow in which the graph is bipartite, each node has unit capacity, and $K=n$.  Yet, there has been no algorithm known for min cost $K$-flow that has complexity matching that of the assignment problem.  Our algorithm for the  min cost $K$-flow and its special case assignment problem has complexity $O((n+K)(m+n\log n))$, which matches the complexity of the fastest known strongly polynomial algorithm for assignment \cite{FT}, but with a very different technique.  It is noted that the algorithm of Fredman and Tarjan \cite{FT} for the assignment problem relies on the bipartite structure of the network and cannot be extended to a non-bipartite network.  One such extension is the unit (vertex) capacity network, in which the minimum cut (or maximum flow) capacity is bounded by $O(n)$.  Our AMC algorithm solves the min cost unit capacity network flow problem in the same run time as the assignment problem.  The only other specialized algorithms to date, for the unit capacity network flow problem, are non-strongly polynomial algorithms:  For instance, the algorithm of Gabow and Tarjan \cite{GT} for unit vertex capacity networks is of complexity $O(\min\{ m^{\half}, n^{\23} \}m \log(nC))$, where $C$ is the largest cost coefficient.

For unit arcs' capacities networks, the maximum flow through the network is bounded by $O(m)$.  Hence the complexity of AMC algorithm for MCF on unit arcs' capacities is $O(m(m+n\log n))$--a speed-up by a $O(\log n)$ factor compared to $MCF(n,m)$.

For convex cost $K$-flow, the complexity of AMC algorithm is the same as its complexity for the standard, linear costs, min cost $K$-flow,  $O((K+n)(m+n\log n))$.  There are no previously known algorithms that have better complexity for convex cost MCF when the flow amount it bounded.  One of the most efficient, non-strongly polynomial algorithms for convex cost MCF is by \cite{AHO03} of complexity $O(mn\log {\frac{n^2}{m}}\log nC))$ where $C$ is the largest cost coefficient.  It is also possible to utilize the fact that on each arc the flow may not exceed $K$ and hence represent each convex cost function on an arc as a convex piecewise linear function on $K$ pieces by multiplying each arc $K$ times.  The strongly polynomial algorithm of \cite{Orl93} then has the complexity of $O(mK\log n (mK+n\log n))$ for the convex cost $K$-flow.  This is more than an order of magnitude worse than the complexity of AMC algorithm.

AMC algorithm also improves considerably on the complexity of algorithms for several classes of the TCT problem.  The uniform cost TCT problem has identical expediting costs for all arcs.  For a project network on $n'$ activity arcs and $m-n'$ precedence arcs the complexity of AMC algorithm is $O((n+n')(m+n\log n))$, which improves by a factor of $O(\log n)$ on the best complexity known to date, $MCF(n,m)$.

The TCT problem with reward of $K$ units per unit reduction in the project finish time is solved in $O((K+n)(m+n\log n))$ with the AMC algorithm.  This again is an improvement compared to $MCF(n,m)$ for $K$ that is $O(m \log n)$.  This $K$-reward problem is the dual of the min cost $K$-flow problem as shown in Section \ref{sec:primal-dual}.
The AMC algorithm solves convex cost TCT with reward of $K$ per unit reduction in the project finish time in the same time as the linear version of this TCT problem, in $O((K+n)(m+n\log n))$.

Table \ref{tab:1} summarizes some of the complexity results derived here as compared to the fastest strongly polynomial algorithms, other than successive shortest paths, known to date. 

\begin{table} [htb]
\begin{center} 
\begin{tabular}{|l|l|l|l|} \hline
{\bf Problem name} & {\bf Complexity} & {\bf  Reference}  & {\bf Here}
\\ \hline \hline
min cost $K$-flow & $MCF(n,m)$ &  & $O((K+n)(m+n\log n))$ \\
unit vertex capacities MCF & $MCF(n,m)$ &  &   $O(n(m+n\log n))$ \\
assignment problem & $O(n (m+n\log n))$ & \cite{FT}  &  $O(n(m+n\log n))$\\
unit arc capacities MCF & $MCF(n,m)$ &   &   $O(m(m+n\log n))$ \\
convex cost $K$-flow** & $O(mn\log {\frac{n^2}{m}}\log nC))$ & \cite{AHO03} & $O((K+n)(m+n\log n))$ \\
\hline
convex cost TCT, reward $K$** & $O(mn\log {\frac{n^2}{m}}\log nC))$ & \cite{AHO03} & $O((K+n)(m+n\log n))$ \\
TCT reward $K$ & $MCF(n,m)$ & \cite{Ful61,Kel61} &  $O((K+n)(m+n\log n))$\\
TCT uniform cost* & $MCF(n,m)$ &  \cite{Ful61,Kel61} &  $O((n+n')(m+n\log n))$\\
\hline \hline
\end{tabular}
\vspace{0.1in}\caption{A comparison to previous fastest (strongly polynomial, if known) algorithms. Here $MCF(n,m)=O(m\log n (m+n\log n))$, \cite{Orl93}.  *For $n'$ activity arcs and $m-n'$ precedence arcs. ** An alternative, strongly polynomial, algorithm based on Orlin's algorithm \cite{Orl93} runs in $O(mK\log n (mK+n\log n))$.} \label{tab:1}
\vspace{-0.31in}
\end{center}
\end{table}

\subsection{The fundamental approach of the AMC algorithm}
Our AMC algorithm builds on the idea of a repeated cuts algorithm, which is a primal algorithm for the TCT problem.  However, instead of evaluating one minimum cut per each unit reduction of the makespan, as done in Philipps and Dessouky's algorithm, the AMC algorithm computes {\em all} minimum cuts of the same value in $O(m+n\log n)$, given an initial minimum cut. Consequently, the makespan is reduced from its value at one breakpoint to the value at the next breakpoint in $O(m+n\log n)$, plus the complexity required to update the maximum flow (minimum cut) from the previous breakpoint's maximum flow.  The complexity then depends on the number of breakpoints in the TCT curve, a number that cannot exceed the value of the maximum flow in the respective graph.  This value is ``small" for various MCF problems such as MCF on unit capacity networks and the assignment problem.

We comment that the TCT problem setting used to solve MCF problems is more general, and sometimes appears to be non intuitive, as compared to standard TCT problems that come up in applications in project management.  For instance, the activities can have durations that are {\em negative} and can have {\em no lower bounds} even though that does not make intuitive sense in the context of project management. The project network is also permitted to contain cycles, as opposed to the standard DAG project network, provided there are no {\em positive} duration lengths cycles.  The TCT problem is still well defined for this generalization: The earliest finish time of the project is the longest path in the graph from start to finish regardless of whether the weights of the arcs are positive or negative.  Likewise, the expediting (or reducing) of durations will result in reduced project finish time (or longest path), whether positive or negative.

\begin{comment}\label{com:DAG}
In the presentation here we use the acyclic structure of the DAG project network in the use of the critical path method (CPM). This is so to retain the intuitive link of the labels to earliest or latest start times of activities in the presentation.   CPM  is used to find the longest/shortest path labels of nodes, and runs in linear time $O(m)$.   In non-acyclic networks CPM is replaced by Dijkstra's algorithm using a standard technique of utilizing the node labels (potentials) to maintain non-negative reduced arc costs without changing the shortest paths solutions.   This increases the run time of CPM from $O(m)$ to $O(m+n\log n)$ which is still dominated by the overall complexity of the calls to the all-min-cuts iterations and therefore does not affect the overall running time.  See Section \ref{sec:node-potential} for more details.
\end{comment}

\subsection{Overview}
In the next section Section \ref{sec:primal-dual} we prove the primal dual relationship of min cost $K$-flow and TCT with $K$ reward.
In Section \ref{sec:prelim} we provide notation and preliminaries on the problem of time-cost tradeoff problem and basic in project management, including the CPM (Critical Path Method) the concepts of slack and earliest and latest start times. We present also the maximum flow minimum cut problems and the concept of residual graph.
In Section \ref{sec:PD} we present a sketch of Philips and Dessouky's algorithm and the repeated cuts approach for solving the TCT problem.  Section \ref{sec:AMC} contains the details of the AMC algorithm and its major subroutines for finding the bottleneck of a cut, the all-min-cuts stage, and the finding of critical or negative slacks arcs that result from the bottleneck reduction in duration of arcs in a cut.

In Section \ref{sec:solvingTCT} it is shown how to utilize the AMC-algorithm for solving TCT problems.  Section
\ref{sec:MCKF} demonstrates the use of AMC-algorithm for solving the minimum cost $K$ flow problem, whether convex or linear.  In Subsection \ref{sec:assignment} we show a detailed example of the application of AMC-algorithm to an assignment problem.
Section \ref{sec:warmstart} discusses ``warm start" ideas for node potentials (or their earliest/late start times) that can speed up the algorithm in practice.  We also present additional details on how to implement the equivalent of CPM for cyclic graphs.  We conclude with several remarks and potential future research in Section \ref{sec:conclusions}.



%

\section{Primal dual relationship of TCT and MCF} \label{sec:primal-dual}

We present here the primal-dual relationship of TCT and MCF when the dual of TCT is a standard $K$-flow MCF problem.  
Recall that the TCT problem is well defined even for negative durations, and even without lower bounds on feasible durations.  The finish time of the project is still the longest path from $s$ to $f$, which can be computed in linear time on a DAG network.

Consider the formulation of TCT problem with no lower bounds, $\underline{d}_{ij}=-\infty$, and with reward per unit reduction in the project finish time of $K$ units.  It is shown here that this TCT problem is the dual of minimum cost $K$-flow problem. The formulation with the respective notation for the dual variables is:

\[
({\mbox{\sf Primal-TCT}}) \hspace{.1in}\begin{array}{lll}
\mbox{} \max \ &  \sum _{(i,j)\in A} w_{ij}  x_{ij}-Kt_f  \\
\mbox{subject to }\ &  x_{ij} + t_i -t_j \leq 0 \ \ \mbox {for all} \quad (i,j)  \in A  & \quad \quad \quad q_{ij}\\
&x_{ij} \leq d_{ij} \ \mbox {for all} \quad (i,j)  \in A  & \quad \quad \quad z _{ij}\\
& t_s=0.
\end{array}
\]

\[
({\mbox{\sf Dual-TCT'}}) 
\begin{array}{ll}
\mbox{} \min \ &  \sum _{(i,j) \in A} d_{ij} z_{ij}   \\
\mbox{subject to }\ &  \sum _{k|(i,k)\in A} q_{ik}-\sum _{k|(k,j)\in A} q_{kj}=0 \ \ \mbox {for all}\ \  k\in V\\
\ &  q_{ij}+z_{ij}  = w_{ij} \ \ \mbox {for all} \quad (i,j) \in A\\
\ &  \sum _{(i,f)\in A} q_{if}=K \\
& q _{ij} , z_{ij} \geq 0
\ \mbox {for all} \quad (i,j) \in A.
\end{array}
\]

Since $z_{ij}$ is nonnegative and  $z_{ij}=w_{ij}-q_{ij}$ we add a constraint $q_{ij}\leq w_{ij}$ and 
substitute for $z_{ij}$ in the objective function. The resulting objective is equivalent to the function $ \min   \sum _{(i,j) \in A} -d_{ij} q_{ij} $. This simplified formulation is then:
\[
({\mbox{\sf Dual-TCT}}) 
\begin{array}{ll}
\mbox{} \min \ &  \sum _{(i,j) \in A} -d_{ij} q_{ij}   \\
\mbox{subject to }\ &  \sum _{k|(i,k)\in A} q_{ik}-\sum _{k|(k,j)\in A} q_{kj}=0 \ \ \mbox {for all}\ \  k\in V\\
\ &  \sum _{(i,f)\in A} q_{if}=K \\
&0\leq  q _{ij}  \leq w_{ij}
\ \mbox {for all} \quad (i,j) \in A.
\end{array}
\]

This {\sf Dual-TCT} problem is precisely the minimum cost $K$-flow problem, $K$-MCF.  In the figure below the dual of TCT for finite duration lower bounds is contrasted with the absence of lower bounds.

\begin{figure}[h!]
\vspace{-1.10in}
  \begin{center}
 \scalebox{0.5}
  { \hspace{-0.34in}
      \epsfig{figure =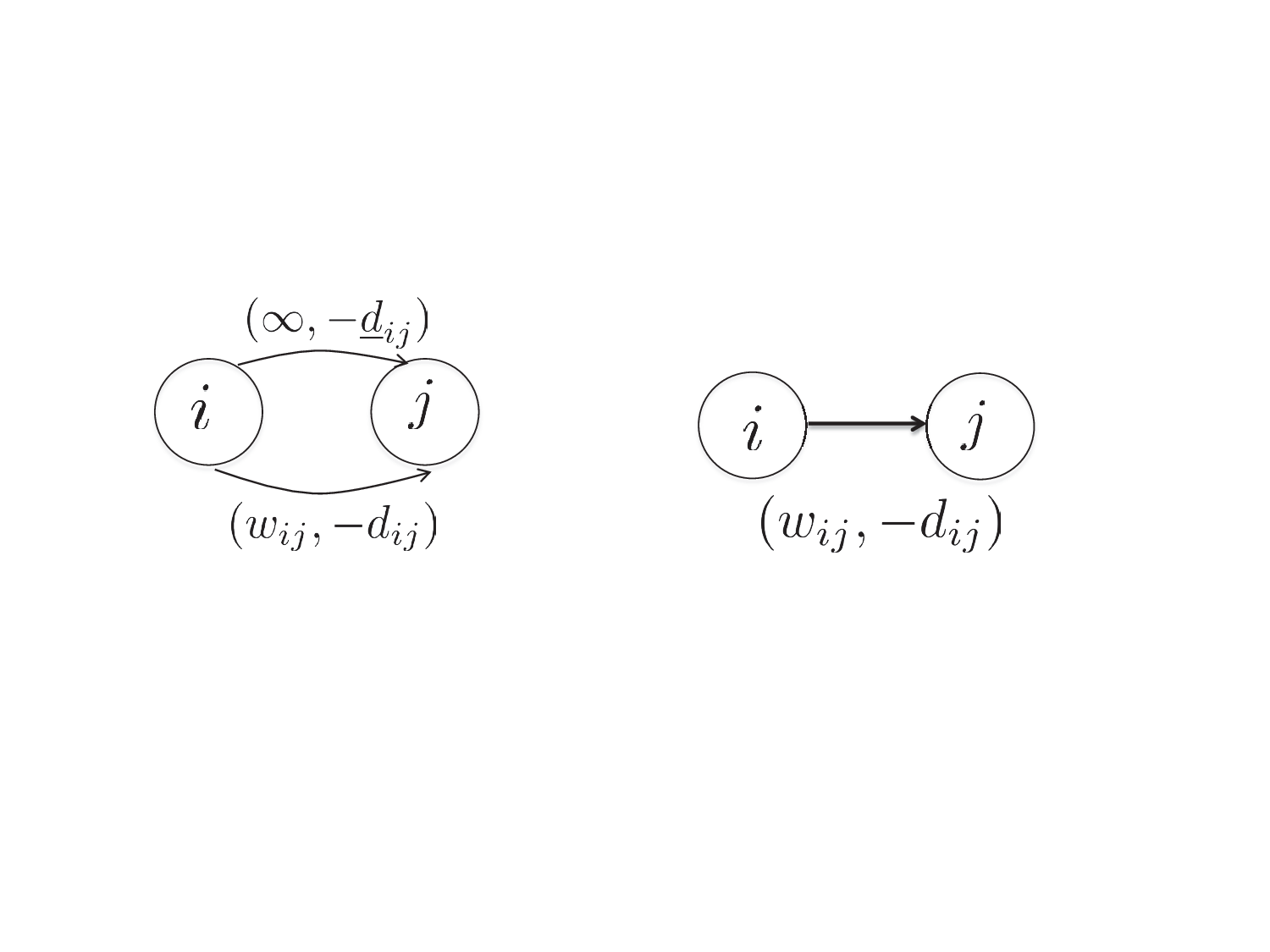}
      }
      \end{center}
\vspace{-1.45in}
  \caption{The arcs for activity $(i,j)$ for general expediting costs $w_{ij}$ and finite duration lower bound, $\underline{d} _{ij}$, and with no duration lower bound. The notation is (upper-bound,cost).}\label{fig:jj'}
\end{figure}

The convex cost $K$-flow problem is the dual of the convex expediting costs TCT, with the convex expediting cost functions $W_{ij}(x_{ij})$ for each activity arc.   The graph representation of the convex cost $K$-flow has, for each arc $(i,j)$, multiple arcs.  Since we are interested in integer solutions, the convex expediting cost function $W_{ij}(x_{ij})$, is represented, {\em implicitly}, as a collection of unit capacity arcs going from $i$ to $j$ where the cost of the arcs are: $w^k_{ij}= W_{ij}(d_{ij}-(k-1))-W_{ij}(d_{ij}-k)$, where $W_{ij}(d_{ij})=0$.  Due to the convexity, $w^k_{ij} \leq w^{(k+1)}_{ij}$.   This representation is implicit in the sense that these multiple arcs are only evaluated as needed during the algorithm.  Similar idea applies for piecewise convex linear functions with up to $K$ pieces, but the length of each linear piece is not necessarily one unit.  In that case the number of arcs is equal to the number of linear pieces and the capacity of each arc is the respective length of the linear piece in the function.

\section{Preliminaries and notations}\label{sec:prelim}

A project network formulated using AOA (Activity-On-Arc) representation is an $s,t$-graph $G=(V,A)$ where node $s$ represents the start of the project and $t$ the finish of the project. (Henceforth we denote the node that indicates the finish of the project by $t$, rather than $f$.)  Each activity has an arc associated with it, the activity arc, and there may be additional arcs needed to represented precedence relations, called precedence arcs.    An activity $(i,j)$ has duration $d_{ij}$ and expediting cost $w_{ij}$.  Precedence arcs have $0$ durations and expediting cost of $\infty$.  Node $s$ has no predecessors, and node $t$ has no successors. An activity $(i,j)$ can only start once all its predecessors arcs activities have been completed.  The project management problem is considered well defined only if the project graph is a directed acyclic graph, a DAG.  The earliest finish time of the project is the longest path from $s$ to $t$, in terms of the sum of the durations of the activities on the path.  The longest path(s) is (are) found, using dynamic programming, in linear time $O(|A|)$.  The longest paths are called critical paths and all arcs and nodes that reside along the longest paths are called {\em critical} arcs and nodes respectively.

In TCT the reduction in the durations of the activities is referred to as crashing or expediting of activities.   Each activity has a {\em normal} duration, and a minimum possible duration.  The normal duration of the project is the earliest finish time of the project with the activities at their normal durations.  The minimum possible duration of the project is the earliest finish time for the activities at their minimum durations.

Let $d_{ij}$ be the normal duration of activity $(i,j)$, $\underline{d} _{ij}$ the minimum possible duration of the activity, and $w_{ij}$ the cost of expediting the activity by one unit.  Dummy activities that correspond to precedence arcs have a fixed normal duration of $0$ and cost per unit expediting of $\infty$.   Feasible activity durations $x_{ij}$ satisfy, $\underline{d} _{ij}\leq x_{ij} \leq d_{ij}$.
For a feasible durations vector $\x$ we let the associated earliest finish time of the project be denoted by $T(\x )$.  This finish time, as well as other quantities, is found with the critical path method CPM described below.

Let $T^{0} = T(\d )$ be the normal project duration, or the earliest project finish time, for activities at their normal durations $d_{ij}$.  Let $T^{min}= T(\underline{\d})$ be the minimum possible project duration corresponding to activity durations at the minimum level $\underline{d} _{ij}$.

{\bf The Critical Path Method CPM and the concept of slacks}:  Given a DAG project network and activity durations, one employs a simple dynamic programming, called CPM, in order to determine critical paths, arcs and nodes and evaluate slacks.    Each node $i$ gets two labels: $ET(i)$ and $LT(i)$, the earliest start time and the latest start time, respectively.  The label $ET(i)$ is the longest path distance from $s$ to node $i$ and is also the earliest time by which activities that originate in $i$ must start in order for the finish time to be earliest possible (the longest path distance).  Once $ET(t)$ is determined, it is set to $T^0$, the earliest finish time of the project.  Applying backwards dynamic programming from node $t$ each node $i$ gets a second label $LT(i)$ that indicates the latest time by which an activity arc originating in $i$ (or ending in $i$) must start (finish) in order for the project to finish at time $T^0$.

For the convenience of the readers we provide the dynamic programming recursions used in the {\bf CPM} (Critical Path Method) to determine the longest paths, the activity slacks and the node labels.  Setting the boundary condition $ET(s)=0$, the recursion proceeds according to the topological order in the DAG (compute the function $ET(j)$ once all its predecessors function values have been computed):
$$ET(j)= \max _{i|(i,j)\in A} \{ ET(i) +d_{ij}\}.$$
The process terminates when $ET(t)$ has been computed. The backwards dynamic programming recursively computes $LT(i)$ for all nodes in reverse topological order. The boundary condition is $LT(t)=ET(t)$:
$$LT(j)= \min _{p|(j,p)\in A} \{ LT(p) -d_{jp}\}.$$

The {\bf total slack} (also known as total float) of an activity $(i,j)$ of duration $d_{ij}$ is $s_{ij}=LT(j)-ET(i)-d_{ij}$.  We refer to the total slack here as {\em slack} without risk of ambiguity as there is no use made here for other types of slacks (free slack).   For an activity durations vector $\x$, let $s_{ij}(\x)=LT(j)-ET(i)-x_{ij}$ be the slack of activity $(i,j)$ where $LT(j)$ is computed with respect to $T(\x )$.   An arc is said to be {\em critical} if it lies on a longest path  (critical path) which occurs if and only if its slack is $0$.  A node $i$ that lies on a critical path satisfies that $ET(i)=LT(i)$.  We'll call such nodes {\em critical nodes} and denote the set of critical nodes by $V_c$.   An alternative characterization of arc criticality is stated in the following claim:


\begin{claim} \label{claim:criticalarc}
An arc $(i,j)$ is critical if and only if its endpoints $i$ and $j$ are critical and its duration satisfies, $d_{ij}=ET(j)-ET(i)=LT(j)-LT(i)$.
\end{claim}
Note that the equality $ET(j)-ET(i)=LT(j)-LT(i)$ applies to all pairs of critical nodes $i,j\in V_c$ since $ET(i)=LT(i)$ and $ET(j)=LT(j)$.  Also note that a non-critical arc {\em can} have both endpoints that are critical nodes. (See e.g.\ in the assignment algorithm example, Figure \ref{fig:final}(A),(B), has arc $(3,2')$ with critical endpoints, but the arc is non-critical.)

\begin{cor}\label{cor:labels2durations}
Given an expedited project network, $G=(V,A)$, with valid node labels $\E$, $\L$, and normal durations $\d$, then
the expedited duration of an activity $(i,j)$, $x_{ij}$, is determined by the labels of its endpoints,
\[
x_{ij} = \left\{ \begin{array}{ll}
      LT(j)-ET(i) & \mbox{ if     }\  LT(j)-ET(i)\leq d_{ij}\\
      d_{ij} & \mbox{ if     }\ LT(j)-ET(i) > d_{ij}.
                \end{array}
\right. \]
\end{cor}

{\bf Adjusted CPM for cyclic graphs}: As noted in the introduction, the project management problem and the TCT problem are well defined even if the network is cyclic, provided there are no positive sum  of durations directed cycles. For the respective MCF, which is the dual of TCT, that means that there are no negative cost cycles in the network.  The labels $ET(j)$ are then the longest paths distances of $j$ from $s$.  The labels $LT(j)$ are computed as the longest paths distances in the reverse graph of $j$ from $t$ and then subtracting these distances from $ET(t)$.

In order to compute the longest paths distances it is convenient to replace the durations of arcs by multiplying them by $-1$ and seeking shortest paths distances in the resulting graph. (That graph happens to be also the min cost flow graph for the dual MCF problem.)  We then apply Bellman-Ford algorithm, once, to compute the shortest paths labels of all nodes from node $s$, and then replace the costs by the reduced costs that are guaranteed to be non-negative.  Once this process has been complete, each computation of the shortest paths labels is done with Dijkstra's algorithm, that requires non-negative costs, in complexity $O(m+n\log n)$, \cite{FT}.


{\bf Max-flow min-cut, cut arcs and residual graphs}:
An $s,t$-cut, $(S,T)$, in a graph $G=(V,A)$ with arc capacities $u_{ij}$ for all $(i,j)\in A$, is a partition of $V$ to two sets $S$ and $T=\bar{S}$ such that $s\in S$ and $t\in T$. 
For a graph with all capacity lower bounds equal to $0$
the {\em capacity of the cut} $(S,T)$  is the sum of the weights on the arcs with tails in $S$ and heads in $T$
defined as: ${C(S,T)=\sum_{(i,j)\in A|i\in S, j\in T}{u_{ij}}}$.  For graphs containing arcs $(i,j)$ in $A$ with positive non-zero lower bounds, $\ell _{ij}$, the capacity of the $s,t$ cut $(S,T)$ is defined as $C(S,T)=\sum_{(i,j)\in A|i\in S, j\in T}{u_{ij}} - \sum_{(i,j)\in A|i\in T, j\in S}{\ell _{ij}}$.  Note that the definition for the $0$ lower bounds is a special case.  For a cut $(S,T)$ the arcs $(i,j)$ such that $i\in S, j\in T$ are referred to as the {\em cut-forward} arcs, and the arcs $(i,j)$ such that $i\in T, j\in S$ 
are referred to as the {\em cut-backward} arcs.

To find a minimum $s,t$-cut in an $s,t$-graph $G=(V,A)$, we find first a maximum flow in the graph.  Let $f_{ij}$ denote the flow value on arc $(i,j)\in A$.  A flow vector $\f =\{f_{ij}\}_{(i,j) \in A}$ is said to be {\em feasible}
if it satisfies
\begin{compenum}
\item [(i)] flow balance constraints: for each $j \in V\backslash \{s,t\}$, $\sum_{(i,j)\in
    A} f_{ij} = \sum_{(j,k)\in A} f_{jk}$ (i.e., inflow($j$) =
    outflow($j$)), and
\item [(ii)] capacity constraints: for all $(i,j)\in A$, the flow value is
    between the lower bound and upper bound capacity of the arc, i.e., $\ell _{ij}
    \leq f_{ij} \leq u_{ij}$.
\end{compenum}
For a flow vector $\f =\{f_{ij}\}_{(i,j) \in A}$ and any cut partition $(S,T)$, $f(S,T) = \sum_{i\in S, j \in T}f_{ij}- \sum_{p\in T, q \in S}f_{pq}$.  The {\em value} of the flow is the amount of flow on any $s,t$-cut, e.g.\  the value, $f(\{ s\},V \setminus \{ s\})$.

%
%
%

Given a feasible flow $\f$, an arc $(i,j)\in A$ is said to be {\em
saturated} if $f_{ij}=u_{ij}$.  An arc $(i,j)$ is said to be a {\em
residual arc} if $(i,j)\in A$ and $f_{ij}<u_{ij}$ or if $(j,i)\in A$
and $f_{ji}>\ell _{ij}$.   For $(i,j)\in A$, the {\em residual capacity} of arc
$(i,j)$ with respect to the flow $\f$ is $u_{ij}^{\f}=u_{ij}-f_{ij}$, and
the {\em residual capacity} of the reverse arc $(j, i)$ is
$u^{\f}_{ji}=f_{ij}-\ell _{ij}$. We denote the residual
capacity of arc $(i, j)$ by $r_{\f}(i,j) = u_{ij}^{\f}$.  If the residual capacity of an arc or reverse arc $(i,j)$ is positive, $r_{\f}(i,j)>0$, then arc $(i,j)$ is said to be a {\em residual arc}

Let $A^{\f}$ denote the
set of residual arcs for flow $\f$ in $G$, and let $G^{\f} = (V, A^{\f})$ denote the {\em residual network with respect to} $\f$.  An augmenting path with respect to a feasible flow $\f$ is a path from $s$ to $t$ in the residual graph $G^{\f}$.  Ford and Fulkerson's max-flow min-cut theorem, \cite{FF56}, states that a flow $\f$ is maximum if and only if the residual graph $G^{\f}$ has no $s,t$ path.  Indeed, if a flow is maximum then a minimum cut of capacity equal to the flow value is attained by labeling all nodes reachable from $s$ in the residual graph.  Since the flow is maximum the set of labeled nodes does not include $t$.  Denoting the set of labeled nodes by $S$, the cut $(S,\bar{S})$ has capacity equal to the value of the flow and is thus a minimum cut (\cite{FF56}).



%

\section{A sketch of Phillips and Dessouky's algorithm} \label{sec:PD}
Phillips and Dessouky's algorithm (PD-algorithm) is based on employing repeated cuts.  This approach relies on the observation that in order to reduce the duration of the project by one unit, all critical paths must have their length (total duration) reduced by one unit.  In a constructed $s,t$ graph containing only the critical arcs, any $s,t$ cut $(S,T)$ is a set of critical arcs so that a unit reduction in duration of each arc in the set results in a unit reduction in the project duration.  The key is to assign capacities to the critical arcs so that a modification to the durations of the arcs in a minimum cut will correspond to a minimum cost unit reduction in the project duration.

At an iteration of PD-algorithm a minimum $s,t$-cut is found in a subgraph induced by the critical arcs. For appropriately defined capacities, if the cut capacity is finite, the duration of the project is reduced by one unit at a cost equal to the cut capacity.  Otherwise the project duration may not be reduced from its current value since it is already at its minimum possible duration, $T^{min}$.  

%
%
%
%

For a vector of normal durations $\d$ and a feasible durations vector $\x$ with the corresponding set of critical activity arcs, $A(\x)$, the respective capacitated {\em critical graph} $G_{\x}= (V,A(\x) )$ is constructed by assigning every critical activity arc $(i,j)\in A(\x)$ capacity lower and upper bounds, $(\ell _{ij} (\x), u_{ij}(\x))$, as follows:

\[ (\ell _{ij}(\x), u_{ij}(\x))= \left\{ \begin{array}{ll}
      (0, w_{ij})  & \mbox{ if     }\  \underline{d} _{ij} <  x_{ij}= d_{ij}\\
      (w_{ij}, w_{ij}) & \mbox{ if     }\ \underline{d} _{ij} <  x_{ij}< d_{ij} \\
      (w_{ij}, \infty) & \mbox{ if     }\ \underline{d} _{ij} =  x_{ij}< d_{ij}\\
      (0, \infty) & \mbox{ if     }\ \underline{d} _{ij} =  x_{ij} = d_{ij}.
                \end{array}
\right. \]

All precedence arcs have capacity $\infty$.  Phillips and Dessouky 1977 \cite{PD77} proved that for a finite $s,t$ cut $(S,T)$ in graph $G_{\x}$, the following update of cut arcs' durations yields a minimum cost unit reduction in the project duration time:

\[
x_{ij} := \left\{ \begin{array}{ll}
      x_{ij} -1 & \mbox{ if     }\  i\in S, j\in T\\
      x_{ij} +1 & \mbox{ if     }\ i\in T, j\in S {\mbox {\ and }} \ell _{ij} = w_{ij}.
                \end{array}
\right. \]

The cost of this one unit reduction in the duration of the project is $\sum _{i \in S, j \in T}w_{ij} - \sum _{j \in T, i \in S} w_{ji}$.  Because this cost is also the capacity of the respective cut, $(S,T)$, then the minimum cost reduction of project duration is achieved for a cut of minimum capacity.  The critical arcs graph is then updated and this minimum cut procedure is repeated for a number of iterations equal to the number of units reduction desired in the project duration.   For the goal of reducing the project finish time for $T^0$ to $T^*$ the complexity of the algorithm is therefore dominated by $T^0-T^*$ calls to a minimum cut procedure.  A variant of this algorithm was shown in \cite{Hoc15} to run in polynomial time with $n\log {\frac{|T^0-T^*|}{n}}$ calls to a minimum cut procedure.

\section{The AMC (all-min-cuts) algorithm}\label{sec:AMC}
AMC algorithm is a repeated cuts algorithm for solving the TCT problem which also generates the respective TCT curve.  This implies that the algorithm provides implicitly also an optimal solution for {\em any} project makespan in the range of the TCT curve, yet its complexity does not depend on the number of values in the range.  As compared to the PD-algorithm, AMC algorithm differs in a number of ways:

\begin{compenum}
\item It is using a {\em crashing graph} induced on critical nodes instead of the critical graph induced on critical arcs used in PD-algorithm.
\item Bottleneck reduction: Instead of one unit reduction as in PD-algorithm, AMC algorithm reduces the project duration, for each cut, by a maximum amount possible, referred to as the {\em bottleneck amount}. 
\item All-min-cuts: Rather than finding one cut at a time, the all-minimum-cuts procedure finds all minimum cuts of the same value in the crashing graph. The complexity of the procedure, once a maximum flow is given, is linear time $O(m)$ plus the complexity of finding the bottleneck value and arc(s) in each cut for up to $n$ cuts, which is $O(m+n\log n)$.
\item Our method utilizes the fact that the maximum flow at the end of one iteration that generates all the minimum cuts of capacity equal to the maximum flow value, remains feasible for the next iteration's crashing graph, (see Theorem \ref{thm:properties}).
\item Once a minimum cut is identified, the duration modification by the bottleneck amount guarantees that at least one additional node is reachable from $s$ (proved in Theorem \ref{thm:properties}). This bounds the number of cuts in an iteration by $n$, the number of nodes.
\item AMC algorithm proceeds with finding all-min-cuts of equal value, and {\em delays} the check for created critical, or negative slacks, arcs, until all-min-cuts have been identified.  This is in contrast to the PD-algorithm which employs CPM to check for created critical arcs after each reduction of one unit in the project makespan.
\end{compenum}

We present the features of the AMC algorithm and then present it fully with its subroutines.  First we consider the incorporation of bottleneck reduction and crashing graph in a repeated cuts algorithm; then the all-min-cuts method that finds all minimum cuts of the same value (capacity) and their respective bottlenecks; then the delayed check for critical and negative slacks arcs, that may possibly require adjustment in the bottleneck amounts. Finally the update of the crashing graph for the next iteration that requires to identify newly created critical arcs (done with CPM, or Dijkstra's for cyclic networks).

\subsection{Bottleneck reduction and the crashing graph.} \label{sec:crashing}
An obvious improvement on the unit reduction in the project duration derived with PD-algorithm is to modify the durations of arcs on the cut by the maximum possible amount, which we call the {\em bottleneck value} of the cut.

Our algorithm works with the crashing graph, that contains, not only the critical arcs, but also all the arcs that have critical nodes as endpoints.  For a vector of durations $\x$ let the {\em crashing graph}  be the graph $G_c(\x)= (V_c,A_c)$ induced on the set of critical nodes with respect to $\x$, $V_c=V_c(\x)$.  All critical arcs (w.r.t.\ $\x$) are included in $A_c$, but there can also be a non-empty set of non-critical arcs that have both endpoints in $V_c$, $A_c^+ \subset A_c$. For arcs $(i,j)\in A_c^+ $ the slacks are strictly positive, $s_{ij}(\x ) \geq 1$.  All other arcs in the crashing graph are critical and their slacks are $0$.  This set of critical arcs is denoted by $A_c^0$.  Thus $A_c = A_c^0 \cup  A_c^+ $.  Each of the other non-critical arcs, in $A_{nc}=A\setminus A_c$, has at least one non-critical node as an endpoint.

In the crashing graph $G_c(\x)= (V_c,A_c)$ we set lower and upper bounds
on capacities that correspond to a right subgradient and left subgradient of the respective expediting cost.  Specifically, let $c_{ij}(x_{ij})$ be the duration cost function of arc $(i,j)$:
\vspace{-0.081in}
\[ c_{ij}(x_{ij}) =  \left\{ \begin{array}{ll}
      0 & \mbox{ if     }\ x_{ij} \geq d_{ij} \\
      w_{ij} (d_{ij} -x_{ij}) & \mbox{ if     }\ \underline{d} _{ij} \leq  x_{ij} \leq d_{ij}\\
     \infty & \mbox{ if     }\ x_{ij}< \underline{d} _{ij}.
                \end{array}
\right. \]

It is noted that $c_{ij}(x_{ij})$ can also be any convex function in the interval $[\underline{d} _{ij},{d} _{ij}]$.  Let $c_{ij}^+(x_{ij})=c_{ij}(x_{ij})-c_{ij}(x_{ij}+h)$ be the (absolute value of the) right subgradient of  $c_{ij}()$ at $x_{ij}$ for $h$ a small enough value.  (For the integer durations problem $h=1$.)  Similarly, $c_{ij}^-(x_{ij})=c_{ij}(x_{ij}-h)-c_{ij}(x_{ij})$ is the left subgradient of  $c_{ij}()$ at $x_{ij}$.  Then, the lower bound and upper bound pair, for each arc $(i,j)$ in the crashing graph is $(c_{ij}^+(x_{ij}), c_{ij}^-(x_{ij}))$.  For the linear cost functions $c_{ij}^+(x_{ij})= c_{ij}^-(x_{ij})=w_{ij}$.  We give the details on the adaptation of the algorithm for convex functions in Section \ref{sec:convex}. To simplify the description here we proceed with linear cost functions.
With the definition of the subgradient, the capacity bounds are set as follows:

\[ (\ell _{ij}(\x), u_{ij}(\x))= \left\{ \begin{array}{ll}
     (0, 0)  & \mbox{ if     }\   x_{ij}= d_{ij} \ {\rm and }\  s_{ij}(\x) \geq 1\\
      (0, w_{ij})  & \mbox{ if     }\  \underline{d} _{ij} <  x_{ij}= d_{ij}\  {\rm and }\  s_{ij}(\x)=0\\
      (w_{ij}, w_{ij}) & \mbox{ if     }\ \underline{d} _{ij} <  x_{ij}< d_{ij} \\
      (w_{ij}, \infty) & \mbox{ if     }\ \underline{d} _{ij} =  x_{ij}< d_{ij}\\
      (0, \infty) & \mbox{ if     }\ \underline{d} _{ij} =  x_{ij} = d_{ij}.
                \end{array}
\right. \]

The {\em bottleneck value} of a cut $(S,T)=(S(\x),T(\x))$ is
the minimum among all the feasible reductions/increases in duration that retain the same cost value, among all arcs in the cut.  The bottleneck value is $\delta$ where, 

\[ \delta = \min \left\{ \begin{array}{ll}
      x_{ij} -\underline{d}_{ij}& \mbox{ if     }\ ( i,j)\in (S,T),\ (i,j)\in A_{c}^0\\
      d_{ij} -x_{ij} & \mbox{ if     }\ ( i,j)\in (T,S) {\mbox {\ and }} \ell _{ij} = w_{ij}\\
      s_{ij} & \mbox{ if     }\ ( i,j)\in (S,T),\ (i,j)\in A_{c}^+.
                \end{array}
\right. \]

The update in the durations of the critical arcs of $A_c^0$ on the cut is to adjust them by $\delta$:
\[
x_{ij} := \left\{ \begin{array}{ll}
      x_{ij} -\delta & \mbox{ if     }\  i\in S, j\in T,\ (i,j)\in A_{c}^0\\
      x_{ij} +\delta & \mbox{ if     }\ i\in T, j\in S {\mbox {\ and }} \ell _{ij} = w_{ij}.
                \end{array}
\right. \]

Also, the slacks of $A_c^+$ arcs and cut-backward $A_c$ arcs (at their normal durations) are adjusted due to change in durations:
\[
s_{ij} := \left\{ \begin{array}{ll}
      s_{ij} -\delta & \mbox{ if     }\  i\in S, j\in T, \ (i,j)\in A_c^+\\
      s_{ij} +\delta & \mbox{ if     }\ i\in T, j\in S \  {\rm and }\  \ell_{ij}=0.
                \end{array}
\right. \]

In case the bottleneck amount is $\delta >1$, it is possible that non-critical arcs in $A_{nc}=A\setminus A_c$, outside the crashing graph, become critical, and possibly even assume negative slacks of absolute values up to $\delta -1$.  To see why the negative slacks value cannot be larger than $\delta -1$, note that all non-critical arcs have slacks that are at least $1$, and therefore a change by $1$ unit can create a critical arc, but not a negative slack. To correct for the creation of negative slacks it might be necessary to update the node labels and slacks in the graph induced by $A_{nc}$ after the bottleneck update, and if a negative slack is found, to reduce the duration reduction amount so as to retain non-negativity of slacks.  An important feature of our algorithm is to {\em delay} that step until all minimum cuts of the same value (cost) have been identified.

\subsection{Properties of repeated cuts with bottleneck reduction}
Let $G_c=G_c(\x _1)$ be a crashing graph for a durations vector $\x_1$; $f_{\x _1}$ a maximum flow in $G_c$; and  $(S(\x_1),T(\x_1))$ a respective minimum cut.  Consider a generic step of a repeated cuts algorithm that
begins with activity durations $\x_1$ and then updates the durations of cut arcs in $(S,T)=(S(\x_1),T(\x_1))$ by a quantity $\delta '$ so that $1\leq \delta '\leq \delta $ where $\delta$ is the bottleneck value of the cut. This update results in a durations vector $\x _2$ and a reduced finish time of the project by $\delta '$ units.

\begin{thm}\label{thm:properties}
$\ $ \\
(a) The sets of critical nodes with respect to durations vector $\x _1$ and $\x _2$ satisfy, $V_c(\x _1)\subseteq V_c(\x _2)$.\\
(b) The flow $\f = \f_{\x _1}$ is feasible for crashing graph $G_c(\x _2)$.\\
(c) For $\delta '=\delta$, the bottleneck value, the set of nodes reachable from $s$, in the residual graph $G^{\f}_c(\x _2)$ w.r.t. $\f$ strictly contains $S=S(\x_1)$.
\end{thm}
\noindent
{\bf Proof:}$\ $ \\
(a) Consider the CPM forward dynamic programming.  Any critical path consists of a path from $s$ to a node $u$, followed by arc $(u,v)$ such that $(u,v)$ is in the cut $(S,T)$, followed by a path from $v$ to $t$.   All the nodes in the critical path section up to $u$ belong to $S$, and all the others belong to $T$. Since there are no changes in arc durations for arcs with both endpoints in $S$, then for all nodes $v\in S$:
$$ET(v)_{\x _2}=ET(v)_{\x _1}.$$
Since the lengths of all arcs in $(S,T)$ are reduced by $\delta '$ then, for all nodes $v\in T$:
$$ET(v)_{\x _2}=ET(v)_{\x _1}-\delta '.$$
Specifically, for node $t$, $ET(t)_{\x _2}=ET(t)_{\x _1}-\delta '$.  The CPM backward dynamic programming results then in $LT(v)_{\x _2}=LT(v)_{\x _1}-\delta '$ for all $v\in T$ and $LT(u)_{\x _2}=LT(u)_{\x _1}$ for all $u\in S$.
Consequently, for all $u\in S$ $$ET(u)_{\x _2}=ET(u)_{\x _1}=LT(u)_{\x _2}=LT(u)_{\x _1}$$
thus all nodes in $S$ remain critical.
For all nodes $v\in T$:
$$ET(v)_{\x _2}=ET(v)_{\x _1}-\delta '=
LT(v)_{\x _1}-\delta '=LT(v)_{\x _2},$$
and hence all nodes in $T$ remain critical.
Therefore, all the nodes of the crashing graph $V_c(\x _1)$ remain critical after an iteration.\\
(b) As a result of duration modifications on arc cuts, forward capacities are either unchanged or go up.  Cut-backward arcs are either unchanged or have their lower capacity bound go down to $0$.  In either case, the current flow remains feasible.\\
(c) Let a bottleneck arc be a cut arc (possibly more than one, in case of ties) for which the bottleneck value $\delta$ is attained.  If a bottleneck arc is a cut-forward arc then it is either a critical arc of $A_c\setminus A_c ^+$ now reduced to its minimum duration, and the modified capacity upper bound of that arc is then $\infty$.  Or, the cut-forward bottleneck arc $(i,j)$ is an arc of $A_c^+$, in which case its capacity upper bound is modified from $0$ to $w_{ij}$.  Finally, if a bottleneck arc is a cut-backward arc $(p,q)$ then its lower bound is modified from $w_{pq}$ to $0$.  In all these cases, the modified capacity of the bottleneck arcs makes them forward-residual from a node of $S$ to a node of $T$. And thus at least one additional node in $T$ is reachable in the residual graph for $G^{\f}_c(\x _2)$.

\qed

We remark that the statement in \ref{thm:properties}(b) for $\delta '=1$, that the maximum flow remains feasible after an iteration of PD-algorithm (in which the change is by $1$ unit), was first communicated to the author by Dessouky, \cite{Des14}, and proved in \cite{Hoc15}.

\subsection{All-min-cuts}
A shortcoming of the repeated cuts algorithm is that it may produce multiple minimum cuts, in several consecutive iterations, that are all of equal cost.  Indeed, it is possible that the number of iterations of the PD-algorithm between two breakpoints of the TCT curve is exponential.  With the use of the bottleneck value of the cut, this number is guaranteed to be at most $n$.  This follows from Theorem \ref{thm:properties}(c), as at each subsequent cut at least one node is added to the source set of the cut.   The all-min-cuts algorithm improves on this further: instead of constructing each of the minimum cuts between two breakpoints, all of these minimum cuts are generated with small additional overhead on the complexity of finding the first cut.

It is possible to have instances of TCT for which the number of breakpoints in exponential.  For example, Skutella constructed, \cite{Sku98}, a family of project networks for which the TCT curve has a number of breakpoints which is exponential in the size of the network. 

An all-min-cuts iteration takes as input a maximum flow on the crashing graph that corresponds to the given duration vector, and outputs the sequence of all the $k$ cuts of the same cost as the value of the maximum flow: $(S_1,\bar{S}_1)\ldots ,(S_k,\bar{S}_k)$ where  $\bar{S}_i =V_c \setminus S_i$ and $S_{k+1}=V_c$ is the set of all nodes in the crashing graph.   The following are corollaries of Theorem \ref{thm:properties}(c):
\begin{cor}\label{cor:nested}
The source sets of the cuts are {\em nested}, that is, $S_1 \subset S_2 \ldots \subset S_k$.
\end{cor}
We label critical nodes that are in the source set of the current cut, during an all-min-cuts iteration, by the label $S$, and nodes of $V_c$ that are in the sink set of the current cut by the label $T$.  Initially all nodes of $V_c$, except $s$, are labeled $T$.
\begin{cor}\label{cor:S2T}
During an all-min-cuts iteration, once a critical node's label changes from $T$ to $S$, it remains labeled $S$ for the remainder of the iteration.
\end{cor}

Corollary \ref{cor:nested} implies that one can avoid listing for the output of the all-min-cuts iteration the source sets of the cuts $S_i$, $i=1, \ldots, k$.  Instead, we list only the incremental sets of nodes added at each subsequent cut $\Delta _i=S_i \setminus S_{i-1}$ for $i=1, \ldots, k+1$, where $S_{k+1}=V_c \setminus S_k$ and $S_0=\{ s\}$.  This list is a partition of the set of nodes $V_c$ and its size is at most $n$.



A stage, or an iteration, is a call to the all-min-cuts procedure for one particular value of the flow (and cut). The stage begins with a crashing graph and a maximum flow of value $v$ in the crashing graph.  The bottleneck value of the associated cut is evaluated, and the durations of cut arcs are (implicitly) updated.
According to Theorem \ref{thm:properties} (b), once the durations and slacks are updated for bottleneck arcs, then the capacities in the crashing graph are updated as well, and the existing flow is still feasible for the modified durations and the respective modified capacities crashing graph.  In the new residual graph, with respect to the modified capacities, at least one additional node of $V_c$ is reachable from the source, as stated in Theorem \ref{thm:properties} (c).  The algorithm {\em continues} to find additional nodes reachable from $s$. 
If node $t$ is reachable, then the current flow is no longer maximum and can be augmented.  In that case the stage terminates.  If however $t$ is not reachable, then a new cut of value $v$ is found, with a source set strictly containing the source set of the previous cut and the all-min-cuts stage continues.


%


We describe here the all-min-cuts (AMC) algorithm with the termination rule of not exceeding $K$ (the reward value), as the largest cost per unit reduction in project finish time.  Other termination rules, such as, $K$ equal to the value of the maximum flow in the network, or target project finish time for TCT, are possible as well.  The pseudocode for the algorithm is displayed in Figure \ref{fig:AMC}

\begin{figure}[!!!hbt]
\noindent \hspace{-3.08in}
\underline{\sc all-min-cuts-algorithm} ($\x=\d$, $G=(V,A)$, $K$)
\begin{small}
 \begin{description}
  \item[Step 0]  Compute (with CPM*) critical nodes $V_c(\x )$, values of $ET(j)$ for all $j\in V_c(\x )$.    
 Let $G_c$ be the crashing graph $G_c({\x })=(V_c(\x ), A_c(\x))=(V_c,A_c)$, and  $\f = \0$ a feasible flow vector in $G_c$.
 \item[Step 1] Find max flow $\f ^r$ in the residual graph $G_c^{\f}$. Set $\f \leftarrow \f + \f ^r$ \{the maximum flow in $G_c$\}.
 \item[Step 2]   If flow value $|\f |\geq K$ terminate; output $(\f,\E ,\L)$, else, continue.
\item[Step 3]  Call {\sf all-min-cuts-stage} $(\E ,G_c,\f )$.  Return ($G_c$,$k$,$\Delta _1,\ldots ,\Delta _k, \Delta _{k+1}$; $\delta_1,\ldots ,\delta_k$).
\item[Step 4] If $V_{nc}= \emptyset$, set $p=k$ and $\tilde{\delta}_{k}=\delta _k$ go to step 5;\\
\hspace{-0.11in}Else, \{$V_{nc}\neq \emptyset$\}, call {\sf find-critical}($k$,$\Delta _1,\ldots ,\Delta _k, \Delta _{k+1}$; $\delta_1,\ldots ,\delta_k$); return ($p$,$\delta_1,\ldots ,\tilde{\delta}_{p}$); go to step 5.
\item[Step 5]  \{terminate stage\} Update $\E$ for all $i\in V_c$:\\
 For $i\in \Delta _{\ell}$, ${\ell}=2,\ldots ,p$, $ET(i)\leftarrow ET(i)-\sum _{{j}=1}^{\ell-1}\delta _j$;\\
 for $i\in \cup _{\ell =p+1}^{k+1} \Delta _{\ell}$
 $ET(i)\leftarrow ET(i)-\sum _{{j}=1}^{p-1}\delta _j -\tilde{\delta}_{p}$.  \\
\{note: no change in $ET()$ values for nodes in $\Delta _1$.\}, go to step 1.\\

\end{description}
 \end{small}
 \caption{AMC algorithm. * CPM is replaced by Dijkstra's algorithm for networks that contain cycles.}\label{fig:AMC}
\end{figure}

The AMC algorithm employs two major subroutines.  One is the {\sf all-min-cuts-stage} which takes as input the current maximum flow in the residual graph, and outputs all minimum cuts of capacity equal to the value of the maximum flow.  The second subroutine is {\sf find-critical} which either truncates the sequence of cuts found until at least one non-critical node becomes critical or determines that all min cuts are valid and the flow in the current critical graph can be strictly increased.  We demonstrate later that such truncation can occur at most $n$ times, when $n$ is the number of nodes.

Assuming the validity of the two subroutines, the AMC algorithm is correct since it performs the same task as a repeated cuts with bottleneck reduction algorithm, except that it exploits property Theorem \ref{thm:properties}(b) to find the new cut by using the flow in the previous crashing graph, instead of identifying each minimum cut by finding a maximum flow from scratch.

In the next sections we describe the two subroutines,  {\sf all-min-cuts-stage} and {\sf find-critical}, used by the AMC algorithm

\section{The all-min-cuts stage}\label{sec:all-min-cuts-complexity}
A stage of the AMC algorithm begins with a given maximum flow in $G_c$. The first cut's source set is the set of nodes reachable from $\Delta _0 =\{s\}$.
This source set is $S_1$, the cut is $(S_1,\bar{S}_1)$ and $\Delta _1 =S_1 \setminus \{ s\}$.  At iteration $k$. if $\Delta _k$ does not contain $t$ the next step is to find the bottleneck value of the cut $(S_k,\bar{S}_k)$,  $\delta _k1$, and to update the residual capacities of bottleneck arcs.  This process of finding the nodes reachable from $\Delta _k$ and updating the residual capacities on the bottleneck cut-arcs is repeated until node $t$ is reachable.

After the bottleneck $\delta _q$ is determined for cut $(S_q,\bar{S}_q)$, instead of updating explicitly the durations, or slacks, of arcs on the cut, the bottleneck value is recorded as associated with the nodes of $\Delta _q$:  This recorded information is sufficient to recover later the durations of all arcs as per Corollary \ref{cor:labels2durations}: For all nodes $v\in \Delta _{q-1}$ (which are critical), their $ET(v)=LT(v)$ values are updated $ET(v)\leftarrow ET(v)-\sum _{i=1}^{q-1} \delta_i $.  This allows to reconstruct the duration of an arc $(u,v)$ as $LT(v)-ET(u)$.  If this quantity is greater than $d_{uv}$ then the arc is a non-critical arc of $A_c^+$ in the crashing graph and its slack is equal to $LT(v)-ET(u)-d_{uv}$.

Notations used in {\sc all-min-cuts-stage} include $T$ for the set of nodes in the sink set of the current cut.  For nodes $u\in T$, $bot(u)$ contains the bottleneck value with respect to node $u$ which is the minimum residual capacity among cut arcs adjacent to node $u$ in the source set $S=S_k$. The set of (cut) arcs for which $bot(u)$ is attained is denoted by bot-set$(u)$.  The notation bot-set indicates the set of bottleneck arcs returned by the procedure that identifies the bottleneck value and the bottleneck arcs of the current cut.  The all-min-cuts stage, given in Figure \ref{fig:stage}, makes calls to procedure  {\sc bot-update} that finds the bottleneck value of a cut and the respective set of bottleneck arcs. For Step 1 of {\sc all-min-cuts-stage} the nodes reachable from $\Delta _k$ are determined as follows:  For each $v\in S_k$ (only those newly added nodes, $v\in \Delta _k$, or the set $\Delta '$, can have a residual arc to $T$), if there is an arc $(v,u)\in A_c$ such that $ u \notin S_k$ then $u$ is added to $\Delta _{k+1}$ and to $\Delta '$.

\begin{figure}[!!!hbt]
\noindent \hspace{-3.6in}
Procedure {\sc all-min-cuts-stage} ($\E$, $G_c$, $\f$)
\begin{small}
 \begin{description}
 \item[Step 0]
    Let the residual graph be  $G^{\f}_c(\x )$; $\x ^{(0)}=\x$, $k=0$, $S_0(\x)= \{ s\}$. Let $T= V_c \setminus \{ s\}$; \\ for all $v\in T$ set $bot(v)=\infty$.
\item[Step 1]   Find $\Delta _{k+1}$, the set of nodes reachable from $\Delta _{k}$ in the residual graph $G^{\f}_c$. \\
     \taba If $t\in \Delta _{k+1}$,  stop, output ($G_c$,$k$,$\Delta_1,\ldots ,\Delta_k$;$\delta_1,\ldots ,\delta_k$). \\
    \taba Else, \{For $S_{k+1}= \cup _ {\ell = 0}^{k+1}\Delta _{\ell}$, $(S_{k+1},\bar{S}_{k+1})$ is the respective minimum cut\} set $k:= k+1$.
 \item[Step 2]
 Call procedure {\sc bot-update}($\Delta _k$, $\E$, $G_c$);  return $\delta$ and bot-set.\\
 Let the {\em bottleneck value} of cut $(S_{k},\bar{S}_{k})$ be $\delta _{k}=\delta$.
 \item[Step 3]
Update the residual capacities $(\ell _{ij} (\x ^{(k)}), u_{ij}(\x ^{(k)}))$ for all arcs in bot-set. \{Only the bottleneck arcs can have their residual capacities changed.\}.   Go to step 1.
\end{description}
 \end{small}
 \caption{The all-min-cuts stage.}\label{fig:stage}
\end{figure}

Procedure  {\sc bot-update} is to find the bottleneck value of a cut and set of bottleneck arcs.  It is described next. Recall that the bottleneck for cut $(S_q,\bar{S}_q)=(S,T)$ is defined as,

\[ \delta = \min \left\{ \begin{array}{ll}
      x_{ij} -\underline{d}_{ij}& \mbox{ if     }\ ( i,j)\in (S,T),\ (i,j)\in A_{c}^0\\
      d_{ij} -x_{ij} & \mbox{ if     }\ ( i,j)\in (T,S) {\mbox {\ and }} \ell _{ij} = w_{ij}\\
      s_{ij} & \mbox{ if     }\ ( i,j)\in (S,T),\ (i,j)\in A_{c}^+.
                \end{array}
\right. \]

Although the set of cut arcs changes from one iteration to the subsequent iteration, the cut arcs that are not bottleneck arcs remain cut arcs for the next iteration.  To identify the arcs that assume the minimum, we maintain a sorted list, associated with {\em nodes} of $V_c$.  For each node $v$ in $T$ (a node in the sink set for the current cut), we retain the value $in(v)$ and $out(v)$, where $in(v)$ is the minimum bottleneck value for arcs into $v$, and $out(v)$ is the minimum bottleneck value for arcs out of $v$.

\[  in(v)=\min \left\{ \begin{array}{ll}
 x_{uv} -\underline{d}_{uv}& \mbox{ if     }\  u \in S, (u,v)\in A_{c}^0 \\
 s_{ij} & \mbox{ if     }\   u \in S, (u,v)\in A_{c}^+\\
  \infty &  \mbox{ otherwise} .
                 \end{array}
\right. \]
and

\[  out(v)=\min \left\{ \begin{array}{ll}
 d_{vu} -x_{vu} & \mbox{ if     }\ u \in S, {\mbox {\ and }} d_{vu} > x_{vu} \\
 \infty &  \mbox{ otherwise}.
                 \end{array}
\right. \]


Recall that the vector of durations $\x$ are maintained implicitly via the $\E$ values of each critical node (equal to the $\L$ values for critical nodes).

The set of out-neighbors of  node $i \in V_c$ is denoted by $N^+(i)=\{ j| (i,j)\in A_c\}$, and the set of in-neighbors of $i$ by $N^-(i)=\{ j| (j,i)\in A_c\}$.
Procedure {\sc bot-update}, in Figure \ref{fig:bot-update}, maintains the sorted list of the nodes $u\in T$ according to the key, $bot(u)$, which is their bottleneck value equal to the minimum of $in(u)$ and $out(u)$ for arcs that have the other endpoint in the source set.  This list is maintained as a Fibonacci heap, as discussed below.

\begin{figure}[!!!hbt]
\noindent \hspace{-3.6in}
Procedure {\sc bot-update} ($\Delta _k$, $\E$, $G_c$)
\begin{small}
 \begin{description}
 \item[Step 0] delete nodes of $\Delta _k$ from the sorted list of nodes of $T$;  set $\Delta '=\Delta _k$.
\item[Step 1]  If $\Delta '= \emptyset$, go to step 5; else, let $v\in \Delta '$, $N^+=N^+(v) \cap T$, $N^-=N^-(v) \cap T$.
\item[Step 2]
 If $N^+= \emptyset$, go to step 3; else, select $u \in N^+$; set,
     \[
in(u) := \min \left\{ \begin{array}{ll}
       ET(u)+ET(v)-\underline{d}_{vu} & \mbox{ if     }\ ET(u)-ET(v) \leq d_{vu}\\
      s_{vu}=ET(u)-ET(v)- d_{vu} & \mbox{ if     }\ ET(u)-ET(v) > d_{vu}.
                \end{array}
\right. \]
    \taba If $in(u)<bot(u)$, set $bot(u)\leftarrow in (u)$; bot-set$(u)=\{ (v,u)\}$; ``decrease-key" of $u$ in the sorted list of nodes in $T$.\\
    \taba If $in(u)=bot(u)$, set  bot-set$(u)\leftarrow$ bot-set$(u) \cup \{ (v,u)\}$\\
    \taba $N^+ \leftarrow N^+\setminus \{ u\}$.\\
    Go to step 2.
\item[Step 3]
  If $N^-= \emptyset$, go to step 4; else, select $u \in N^-$; set
     \[
out(u) := \min \left\{ \begin{array}{ll}
      d_{uv} - ET(v)+ET(u)  & \mbox{ if     }\ ET(v)-ET(u) < d_{uv}\\
           \infty &  \mbox{otherwise}.
                \end{array}
\right. \]
    \taba If $out(u)<bot(u)$, set $bot(u)\leftarrow out(u)$;  bot-set$(u)=\{ (u,v)\}$; ``decrease-key" of $u$ in the sorted list of nodes in $T$.\\
    \taba If $out(u)=bot(u)$, set  bot-set$(u)\leftarrow $bot-set$(u) \cup \{ (u,v)\}$\\
    \taba $N^- \leftarrow N^-\setminus \{ u\}$.\\
    Go to step 3.
\item[Step 4]  $\Delta ' \leftarrow \Delta ' \setminus \{ v\}$; go to step 1.
\item[Step 5]  \{terminate\} Return, ``min-key" in $T$: The value $\delta _k= bot(u*)= \min _{u \in T} bot(u)$ and respective bot-set$(u*)$ \{bot-set$(u*)$ may be the union of the bottleneck sets for several nodes attaining min-key\}; stop.
\end{description}
 \end{small}
 \caption{Procedure  {\sc bot-update} for maintaining and updating bottleneck values.}\label{fig:bot-update}
\end{figure}

\vspace{0.1in}

\noindent
{\bf Complexity analysis of the all-min-cuts stage:}  The total number of operations in all-min-cuts stage, excluding the complexity of {\sc bot-update}, is $O(m)$.  This is since each arc in $A_c$ is considered at most once:  Cut arcs that are in the bottleneck set are no longer considered in later iterations.  Procedure {\sc bot-update} is called for during all-min-cuts stage, each time with a different set of nodes $\Delta _k$, and maintains the sorted list of nodes in $T$.  In all these $O(n)$ calls, the procedure scans each arc of $A_c$ at most once.  For each arc $(v,u)$ scanned there can be one update (``decrease-key") in the position of $u \in T$ in the sorted list.  In all calls during all-min-cuts stage there are at most $O(n)$ deletions  (``delete" and ``delete-min") from the sorted list.

The sorted list of nodes of $T$ with respect to the key of $bot(u)$ is maintained with the Fibonacci heaps data structure, \cite{FT}.  Fibonacci heaps data structure requires $O(1)$ operation per ``decrease-key" operation and $O(\log n)$ per ``delete" or ``delete-min". The delete operation takes place only in Step 0, and is executed at most once per node. The total complexity is therefore $O(m+n \log n)$ for all calls made to procedure {\sc bot-update} during an all-min-cuts stage.  This complexity dominates the complexity of the other operations in all-min-cuts stage.  The total complexity of all-min-cuts stage is thus $O(m+n \log n)$.


\subsection{Delayed check for creation of critical arcs.}
It remains to address the procedure {\sc find-critical}.
A naive application of AMC algorithm would be to generate one cut at a time, determine the bottleneck, update the values of $\E$ on the critical nodes (by subtracting the current bottleneck amount from nodes in the sink set) and then check for non-positive created slacks among non-critical arcs of $A_{nc}$ by running CPM (or its Dijkstra's equivalent).  Note that the bottleneck value adjustment of capacities guarantees that no arc of $A_c$ becomes a negative slack arc in the crashing graph.
But a negative slack arc of $A_{nc}$ might be created, in which case the bottleneck value that caused it must be updated so the slack of such arc becomes $0$ (and the arc, along with its endpoints, becomes critical).  If only $0$ slack arcs were created, then no update of the bottleneck value is required.  In either case, when non-positive slack arcs are found, the bottleneck value is adjusted, if there is negative slack, and the all-min-cuts process is terminated with the cut that created new critical arcs.  Then a new crashing graph that contains the newly created critical nodes is generated.  Since whenever a non-positive slack is created, the crashing graph is augmented by at least one critical node, this adjustment of bottleneck value and subsequent update of the crashing graph may happen at most $n$ times.  

The drawback of this naive approach is that it increases the complexity.  At each iteration of all-min-cuts stage, the check for created non-positive slacks require that the $\E$ values of all nodes in the sink set must be updated, and the check for the presence of non-positive slack arcs in $A_{nc}$ requires to traverse at least once all these arcs, at complexity $O(m)$.  Since each run of all-min-cuts stage may generate up to $O(n)$ cuts, the resulting added complexity due to the checks is at least $O(mn)$ steps for each call to all-min-cuts stage, leading to complexity of at least $O(Kmn)$ for a $K$-flow network.   Yet, all these checks, with the exception of possibly $O(n)$ of them, fail to find non-positive slacks and thus result in no change in the crashing graph, and therefore are mostly wasted.

The delayed check is executed after all-min-cuts stage is complete, having generated a total of $k$ cuts.  If the check process finds non-positive slacks for cut $(S_q,\bar{S}_q)$, where $q$ is the lowest index among the $k$ cuts,  then the sequence is adjusted by discarding all cuts $q+1,\ldots, k$, and modifying the bottleneck value $\delta _q$ in case negative slacks were found ($\delta _q$ is reduced by the absolute value of the most negative slack created).  Such adjustment, or backtracking on the all-min-cuts process,
can happen at most $O(n)$ times, since each results in adding at least one critical node to the crashing graph.

Therefore, there is a check for every breakpoint of the TCT curve, and additional $O(n)$ checks that result in backtracking.  Hence, the number of calls to the delayed checks, for a TCT curve with $K$ breakpoints, is $O(K+n)$.  It is shown that the complexity of each check is at most $O(m)$ resulting in total complexity of $O((K+n)m)$ for all checks.

The delayed check is performed in procedure {\sc find-critical},  which takes as input the partition and bottleneck values generated in the all-min-cuts stage, ($G_c$,$k$,$\Delta_1,\ldots ,\Delta_k$;$\delta_1,\ldots ,\delta_k$).
Procedure {\sc find-critical} either identifies the lowest index cut among the cuts generated in the all-min-cuts-stage that creates non-positive slacks, or determines that no new critical node has been created.  In either case, it returns an updated all-min-cuts partition to $q$ $\Delta$ sets and corresponding $\delta _j$ values for which no negative slack is created.

Suppose the first $p$ cuts have been generated and now we determine that there is another cut as $t$ is not reachable with the continuing flow from the residual arcs on the cut $(S_p,\bar{S}_p)$.  The new cut is $(S_{p+1},\bar{S}_{p+1})$.  Once this cut is added, the start times and end times for all nodes in $\Delta _{p+1}= S_{p+1} \setminus S_p$ will remain unchanged in all subsequent iterations throughout this stage.  For start times and end times at the initialization of a stage denoted by $ET^0()$ and $LT^0()$, the start times and end times of the nodes in $\Delta _p$ change at the end of the stage are as follows:

\begin{lemma}
For every (critical) node $u$ in $\Delta _p$, $ET(u)=LT(u)$, and,
   \[
 \hspace{.24in}\begin{array}{ll}
      ET(u)=& ET^0(u)-\sum _{\ell =1}^{p-1} \delta _{\ell}\\
      LT(u)=& LT^0(u)-\sum _{\ell =1}^{p-1} \delta _{\ell}.
\end{array}
\]
\end{lemma}

We introduce additional notation used in procedure {\sc find-critical}:  The complement graph of the crashing graph $G_c$ is  $G_{nc}=(V,A_{nc})$ the subgraph of $G$ induced on the set of non-critical arcs $A_{nc}(\x)$ with respect to $\x$, that have at least one endpoint in $V_{nc}$ (the set of non-critical nodes).

Node $i$ is said to be {\em back-reachable} from $j$ if there is a directed path in $G_{nc}$ from $i$ to $j$.  The set of in-neighbors of $v$, in $G_{nc}$, is $N^+(v)=\{ (u,v)|(u,v)\in A_{nc}$\}. A node $v$ in $V_{nc}$ is said to be visited if its visited-label is true, $visited(v)=1$. We denote by $par(v)$ the parent of node $v$ in the DFS tree.  We call a path of non critical arcs in $A_{nc}$, $(i_1,i_2,\ldots ,i_{p'})$,  a $(\Delta _q,\Delta _p)$-path if $i_1,i_{p'} \in V_c$ with $i_1\in \Delta _q$ and $i_{p'} \in \Delta _p$, all intermediate nodes of the path are non-critical, $i_2,\ldots, i_{p'-1}\in V_{nc}$, and $p'\geq 3$.

Consider now how slacks of arcs in $A_{nc}$ are modified throughout the stage.   Suppose an arc $(i,j)$ lies on a non-critical $(\Delta _q,\Delta _p)$-path between critical nodes $i_1,i_{p'} \in V_c$, $(i_1,i_2,\ldots ,i_{p'})$, with $i_1\in \Delta _q$ and $i_{p'} \in \Delta _p$.   A change in the slack of an arc $(i_{r}, i_{r +1})$ on the $(\Delta _q,\Delta _p)$-path will occur only if the index $q$ is strictly less than $p$.
To see that observe that for  arc $(i_{r}, i_{r +1})$ with original slack $s_{i_{r}, i_{r +1}}$, the {\em modified slack} of the arc with respect to this path is:\\
$\min \{ s_{i_{r}, i_{r +1}}, s_{i_{r}, i_{r +1}} - \sum_{\ell=1}^{p-1}\delta _{\ell} +\sum_{\ell=1}^{q-1}\delta _{\ell}\}= \min \{  s_{i_{r}, i_{r +1}}, s_{i_{r}, i_{r +1}} - \sum_{\ell=q}^{p-1}\delta _{\ell}\}.$

\begin{cor}  Suppose an arc $(i,j)\in A_{nc}$ is on $N$ non-critical $(\Delta _{q_r},\Delta _{p_r})$-paths in $G_{nc}$ between pairs of critical nodes in $\Delta _{q_r}$ and $\Delta _{p_r}$, for $r=1,\ldots , N$.  Then, the modified slack of arc $(i,j)$ with the updated bottleneck durations of arcs in the crashing graph is,\\
$\min \{ s_{ij}, \min _{r=1,\ldots , N}  s_{ij} - \sum_{\ell=1}^{p_r-1}\delta _{\ell} +\sum_{\ell=1}^{q_r-1}\delta _{\ell}\}$.
\end{cor}

Hence the modified slack of a non-critical arc depends only on the indices of the $\Delta$ sets to which the endpoints of the paths it lies on, belong.

Procedure {\sc find-critical} scans, at iteration $q$ for $q=1,...,k+1$, all non-critical paths that end at nodes of $\Delta _q$. If at least one arc on these paths is found to have modified slack that is non-positive with respect to any of these paths, then the process terminates.   Hence, iteration $q$ is initiated only if no nonpositive slack arc was found in $A_{nc}$ for arcs lying on paths the end at nodes of $\Delta _1 \cup ...\cup \Delta _{p-1}$.  Suppose an arc $(i,j)$ of $A_{nc}$ lies on several paths from nodes of $\Delta _{q_{\ell}}$ for $\ell =1,...,r$, such that $q_1 <q_2 ,...< q_r$, that end at nodes of $\Delta _q$.  Then the slack of the arc is modified only if $q_1<q$ and in that case the {\em modified slack} is $s_{ij} - \sum_{\ell=1}^{q-1}\delta _{\ell} +\sum_{\ell=1}^{q_1-1}\delta _{\ell}$.

In particular, the {\em modified slack} of arc $(i,j)\in A_{nc}$  with respect to the smallest index $q'$ so that a node of $\Delta _{q'}$ is back-reachable from node $i$, can be written as $s_{ij} - \sum_{\ell=1}^{q-1}\delta _{\ell} +\sum_{\ell=1}^{q_1-1}\delta _{\ell}=\tilde{s}_{ij} -\sum_{\ell=1}^{q-1}\delta _{\ell} $ where $\tilde{s}_{ij}=s_{ij} +\sum_{\ell=1}^{q-1}\delta _{\ell}$.  We refer to the quantity $\tilde{s}_{ij}$ as the {\em adjusted slack} of arc $(i,j)\in A_{nc}$.


The advantage of using the adjusted slack is that its value is independent of the index $q$.  During iteration $q$, the back-DFS process that starts from nodes of $\Delta _q$  assigns each node $u$ in the tree (of back-reachable nodes) a  label $\tilde{s}(u)$ which is the lowest value of the adjusted slack of arcs of $A_{nc}$ in the subtree rooted at $u$.  Let $\min (q)$ be the minimum value of $\tilde{s}(u)$ among $u \in \Delta _q$. That is, $\min (q) - \sum_{\ell=1}^{q-1}\delta _{\ell}$ is the smallest value of the modified slack on all arcs that are back reachable from nodes of $\Delta _q$.

If that value is non-positive ,that will trigger a termination of the all-min-cuts stage, but it still remains to identify the most negative value of a modified slack, so as to adjust the value of the bottleneck $\delta _{q-1}$.   Since the existence of a non-positive modified slack on some arcs back-reachable from $\Delta _q$ means that the remaining cut partitions are not valid, $\Delta _q:=\Delta _q \cup \Delta _{q+1}\cup...,\cup \Delta _{k+1}$ are merged as the final sink set.  Next, all nodes in the sink set $\Delta _q$ have had their $\tilde{s}(u)$ evaluated, and the value $\min(q)$ is updated, and the smallest modified slack $-\alpha =\min (q) - \sum_{\ell=1}^{q-1}\delta _{\ell}$ is still non-positive.

To update the bottleneck, $\delta _{q-1}$ is replaced by $\delta _{q-1}-\alpha$. With this updated value this bottleneck creates at least one new critical node and critical arc. With this updated value of the bottleneck, all slacks are non-negative.

At each iteration $q$ a back-DFS is initiated from each node $v$ of $\Delta _{q}\subseteq V_c$. Node $v$ is a root of a tree, and any back-reachable node of $V_c$ is a leaf, as well as any node previously visited and labeled.   At the end of that process each visited node $u$ in $G_{nc}$ will have two {\bf node labels}:
$(\tilde{s}(u), p(u))$ defined as follows:\\
The label $\tilde{s}(u)$ is the minimum value of $\tilde{s}_{ij}$ (the adjusted slack) among all arcs $(i,j)\in A_{nc}$ so that node $i$ is back-reachable from node $u$.  The second label $p(u)$ is the lowest index $\Delta _{p(u)}$ such that critical node $w \in \Delta _{p(u)}$ is back-reachable from $u$.

The two labels of each critical node $u$, are initialized to $\tilde{s}(u)=\infty$ and  $p(u)=p'$ for $u\in \Delta _{p'}$ (since the $\Delta$ sets form a partition of $V_c$, such unique index exists). The labels of each non-critical node $u$ in $V_{nc}$, are initialized to $\tilde{s}(u)=\infty$, and  $p(u)=k+1$.  Each arc in $(i,j)\in A_{nc}$ has its (original) slack label $s_{ij}$ computed at the initialization of the stage for the given durations.  These slack values  $s_{ij}$ are necessarily positive for the non-critical arcs.

The back-DFS search is initiated from each node $v$ in $\Delta _q$.  There is back-tracking if either a node of $V_c$ is visited, in which case it is a leaf, or else previously visited nodes of $V_{nc}$ are reached.
All non-root and non-leaf nodes of the DFS tree are non-critical nodes of $V_{nc}$.

We are now ready to describe formally the procedure:

\vspace{0.2in}
\noindent
{\sc Procedure find-critical}($k$,$\Delta _1,\ldots ,\Delta _k, \Delta _{k+1}$; 
$\delta_1,\ldots ,\delta_k$)
\begin{small}
 \begin{description}
 \item[Step 0]\{Initialize:\}
 For all $u\in V$, $ \tilde {s}(u)=\infty$, for $v\in V_{nc}$, $visited(v)=0$, $p(u):=k+1$;\\
 For all $u \in V_c \cap \Delta_{p}$, $p(u)=p$.
Set $q=1$. \{Since nodes in $\Delta _1$ do not have their $\E$ values modified, they cannot be the endpoint of a non-critical path that will cause a reduction in the slack values\}
\item[Step 1] If $q=k+1$, stop; return ($q-1$,$\delta_1,\ldots ,{\delta}_{q-1}$) \\
 \  Else,  $q:=q+1$; $L=\Delta _q$; $\min (q)=\infty$.
\item[Step 2]
If $L = \emptyset$  do\\
  \taba If $\min (q)-\sum _{\ell =1} ^{q-1} \delta _{\ell}>0$ go to step 1;\\
   \taba Else, \\
   \tabb if $q< k+1$, set $L := L\cup \Delta _{q+1} \cup ... \cup \Delta _{k+1}$, $k=q-1$ go to step 2;\\
    \tabb else $\alpha = \sum _{\ell =1} ^{q-1} \delta _{\ell} -\min (q)$; $\delta _{q-1} := \delta _{q-1} - \alpha$ go to step 1.\\
Else, select $v \in L$.
\item[Step 3] If $N^+(v) = \emptyset$ \{backtrack from $v$\} set $visited (v)=1$;\\
\taba If $par(v)=\emptyset$ \{$v \in V_c$ is the root\} $\min (q) =\{ \min (q), \tilde{s}(v) \}$; $L:= L\setminus \{ v \}$; \\
\taba Else, $v:= par(v)$; go to step 3.\\
Else, select $u \in N^+(v)$.
\item[Step 4]  If $visited (u)=1$ or $u \in V_c$ \{backtrack\}, do\\
\taba $p(v):= \min \{ p(v), p(u)\}$; $\tilde{s}(v):= \min\{\tilde{s}(v) ,  s_{uv} +\sum _{\ell =1}^{p(u)-1} \delta _{\ell}  \}$; if $u \in V_{nc}$ $\tilde{s}(v):= \min\{\tilde{s}(v), \tilde{s}(u)\}$.\\
\ Else, $par(u)=v$;  $v \leftarrow u$; go to step 3.
\end{description}
 \end{small}

In Step 2,  if modified slack of non-positive value is found at iteration $q$, the value of $k$ is changed to $q-1$ so as to trigger the termination rule of $q=k+1$ in Step 1.

When Procedure {\sc find-critical} terminates, either all modified slacks of arcs of $A_{nc}$ are positive, and a next stage can be initiated without a modification of the crashing graph, or else, the crashing graph would change as at least one critical node is added to the set $V_c$.  In that case, the all-min-cuts stage is restarted, but possibly without a change in the value of the maximum flow.


The complexity of {\sc find-critical} is linear, $O(m)$, since the dominant work is in the back-DFS process, in which each arc in $G_{nc}$ is inspected at most twice.

\begin{remark}
The reader may question why the same  {\sc find-critical}  procedure could not have been employed whenever cut $(S_{q+1},\bar{S}_{q+1})$ and set $\Delta _q$ are found during the all-min-cuts stage procedure, rather than wait for the completion of the all-min-cuts stage.  The reason is that during iteration $q$, if there is a node $u\in V_{nc}$, the set of non-critical nodes, in the back-DFS tree for which the lowest index back-reachable set is $\Delta _{q'}$ for $p(u)=q'>q$, the set $\Delta _{q'}$ is not known as of yet.  Rather the set of nodes that would ultimately be $\Delta _{q'}$ are in the sink set of the current cut.  As a result, in later iterations and in particular at iteration index $<q'$, the subtree rooted at node $u$ will be scanned again in order to determine the correct $p(u)$ label. This could result in repeated scanning of the arcs up to $O(n)$ times leading for a complexity of $O(nm)$ instead of $O(m)$.
\end{remark}

\begin{theorem}
The all-min-cuts algorithm is correct.
\end{theorem}
{\bf Proof:} The AMC algorithm is a repeated cuts algorithm.   As compared to PD-algorithm the use of the crashing graph and the bottleneck value aggregates a number of consecutive iterations of PD-algorithm into a single iteration.  If this is followed by a check for non-positive slacks' created arcs, and an adjustment to ensure that all slacks are non-negative, then the two procedures are exactly the same.  Also, the use of the all-min-cuts stage with the delayed check for non-positive slacks' created arcs still renders the same result. Since, at the end of a stage, this test is performed (as procedure {\sc find-critical}) and the durations vector that creates at least one $0$ slack arc is restored.  Therefore the two algorithms are equivalent, and the correctness of all-min-cuts algorithm follows.


\qed

\begin{theorem} \label{thm:complexity-K}
The all-min-cuts algorithm for a TCT problem with cut value at most $K$ is $O((K+n)(m+n\log n))$.
\end{theorem}
{\bf Proof:} The complexity of all-min-cuts stage is $O(m+n\log n)$. Suppose the TCT-curve piecewise linear function have slope values $k_1 < \ldots < k_{\ell} \leq K$. The input to each stage is the maximum flow in the crashing graph.  We solve the maximum flow problem with the augmenting paths algorithm.

The complexity of solving the maximum flow in the initial crashing graph $G_c(\d)$ is $O(k_1m)$, as each unit of flow requires $O(m)$ search for an augmenting path.  The second requires $O((k_2-k_1)m)$ complexity, and the $i$th one requires $O((k_i-k_{i-1})m)$. The total complexity of all maximum flows found, for $k_0=0$ is $O( \sum _{i=1}^{\ell} (k_i-k_{i-1})m)$, which is $O(k_{\ell} m)$.

A stage may be interrupted due to the introduction of new critical nodes. Such interruptions however can occur at most $n$ times as this is the number of nodes that can become critical. Procedure {\sc find-critical} is performed at the end of each stage at complexity $O(m)$.  For stages that are not interrupted the total complexity is $O(\ell m)$ which is at most $O(Km)$.  If a stage is interrupted, then the durations are restored so that there is at least one new critical arc in $G_{nc}$ of slack $0$. This may undo all the reduction in the project finish time, except for $1$ unit.  This is since at the beginning of a stage the slacks of all non-critical arcs is at least $1$.  With interruption, the crashing graph gets updated.  We then {\em continue} with the previous maximum, which is either still maximum, and a sequence of cuts is created in the all-min-cuts stage, or else, it is not maximum, in which case the flow is, the c augmented to the next value of $k_i$.  The additional cost of all interruptions is therefore $O(n(m+n\log n))$.

The total complexity of the algorithm is thus $O((K+n)(m+n\log n))$.

\qed


\section{Solving cases of the TCT problem with AMC algorithm}\label{sec:solvingTCT}
\subsection{Reward or budget of $r$ units}
Consider first the TCT problem with reward of $r$ for each unit reduction in the project finish time.  At optimum, the cost of the last unit reduction is $\leq r$.  Therefore, the TCT curve can have at most $r$ different slopes, of values in $\{ 1,2, \ldots ,r\}$.   Per Theorem \ref{thm:complexity-K} the complexity of the all-min-cuts algorithm for the problem is therefore in $O((r+n)(m+n\log n))$.

If the total budget is $r$ then the only possible cost values are in $\{ 1,2, \ldots ,r\}$.  The TCT with $O(r)$ budget is then solved in $O((r+n)(m+n\log n))$.

\subsection{Uniform costs TCT}
When all expediting costs are uniform, they can be scaled to $1$.   Assume that up to $n'\geq n$ arcs are activity arcs, and thus of finite cost.  The other arcs are precedence arcs, which have infinite expediting costs. In that case, any finite capacity cut in the crashing graph has capacity that takes value in $\{ 1,\ldots , n'\}$.  Hence the TCT curve can have at most $n'$ breakpoints and slopes that take values in $\{ 1,\ldots , n'\}$.  With Theorem \ref{thm:complexity-K} the complexity of the all-min-cuts algorithm is at most $O(n'(m+n\log n))$.

\section{Solving the minimum cost $K$-flow problem with AMC algorithm} \label{sec:MCKF}
Solving the min cost $K$-flow problem with AMC algorithm has one issue that requires further comment.  Since there are no durations' lower bounds, $\underline{d} _{ij} =-\infty$, there can be infinite bottleneck value as explained next:  For a given duration vector $\x$ the capacities in the crashing graph arc $A_c$ are,

\[ (\ell _{ij}(\x), u_{ij}(\x))= \left\{ \begin{array}{ll}
     (0, 0)  & \mbox{ if     }\   x_{ij}= d_{ij} \ {\rm and }\  s_{ij}(\x) \geq 1\\
      (0, w_{ij})  & \mbox{ if     }\   x_{ij}= d_{ij}\  {\rm and }\  s_{ij}(\x)=0\\
      (w_{ij}, w_{ij}) & \mbox{ if     }\  x_{ij}< d_{ij}.
                \end{array}
\right. \]

For a minimum cut $(S,T)=(S(\x),T(\x))$ in the crashing graph the bottleneck value is,

\[ \delta = \min \left\{ \begin{array}{ll}
      \infty  & \mbox{ if     }\ ( i,j)\in (S,T),\ (i,j)\in A_{c}^0\\
      d_{ij} -x_{ij} & \mbox{ if     }\ ( i,j)\in (T,S) {\mbox {\ and }} \ell _{ij} = w_{ij}\\
      s_{ij} & \mbox{ if     }\ ( i,j)\in (S,T),\ (i,j)\in A_{c}^+.
                \end{array}
\right. \]

In case all arcs in the cut are cut-forward arcs of $A_{c}^0$, and hence critical, the bottleneck value is infinite.  We address this as follows: this bottleneck value is set to $M$ (big M), and the all-min-cuts stage is terminated.  That is because there exists a value of $M$ large enough so that at least one non-critical arc of $A_{nc}$ and one non-critical node become critical.  The call for the routine {\sc find-critical} is then to search for the smallest value of $M$ so at least one slack of a non-critical arc of $A_{nc}$  becomes zero.  This is done symbolically in $M$, and in linear time, with two values, one is the coefficient of $M$ and the other is a constant.  The minimum adjusted slack is then ``lexicographic" with the coefficient of $M$ being the dominant.   An illustration of how such a procedure works, for the example of applying the AMC algorithm to the assignment problem, is given in Section \ref{sec:assignment}.

Next we provide details on how the optimal solution to the min cost $K$-flow problem is related to the optimal solution derived by AMC algorithm for the respective dual TCT problem with $K$-reward for unit reduction in the project finish time.
Recall that the AMC algorithm terminates when for a given duration vector $\x$, the maximum flow in $G_c=G_c(\x)$, $\f = \f _{\x}$, has value $|\f |= K$.  

The flow $\f$ has positive values only on critical arcs.  To see that note that all non-critical arcs of the crashing graph, $A_c^+$, have capacity $0$ in the graph $G_c$ where the maximum flow is found.   For each critical arc of $G_c$, $(i,j)\in A_c^0$, the slack is $0$, $LT(j)-ET(i)-x_{ij}=0$. The endpoints of a critical arc $(i,j)$, are critical nodes, $ET(i)=LT(i)$, and in the formulation of  {\sf Primal-TCT} $t_i=ET(i)=LT(i)$.  From complementary slackness conditions it follows that the flow variables $q_{ij}$ in the $K$-flow problem {\sf Dual-TCT} can only be positive when the respective arc $(i,j)$ is critical.

We observe that the slacks are the negatives of the {\em reduced costs} of the respective arcs that are usually denoted by $c_{ij}^\pi = c_{ij} -\pi _i + \pi _j \geq 0$ for $c_{ij}$ the cost of the flow on the arc, and $\pi _i$ the {\em potential} of node $i$.  Here the duration $x_{ij}$ is the negative of the cost of the arc, and the node potentials are the respective $t_i$ values that are well defined for critical nodes.

Since the values of $\x$ and the $t_i$ are optimal for {\sf Primal-TCT}, the feasible $K$-flow on the graph of critical arcs $\f$ is an optimal solution to the $K$-flow problem.  This solution is precisely the flow $\f _{\x}$ of value $K$, that once obtained, triggers the termination of the algorithm.  Therefore the output flow $\f _{\x}$ is the optimal min cost $K$-flow.

\subsection{Min Convex cost $K$-flow problem}\label{sec:convex}
\begin{figure}[h!]
\vspace{-0.07in}
  \begin{center}
 \scalebox{0.51}
  { \hspace{-1.2in}
      \epsfig{figure =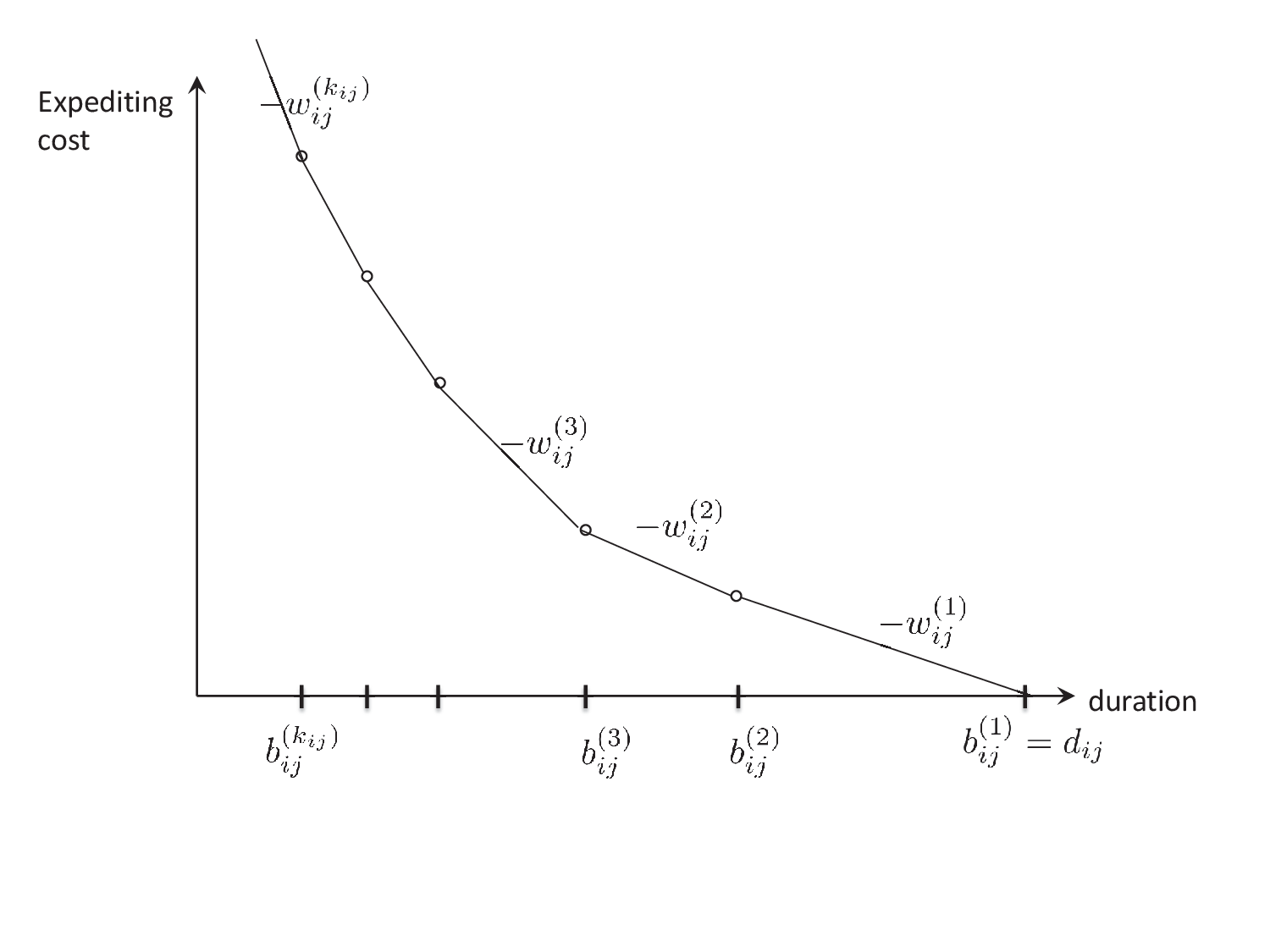}
      }
      \end{center}
\vspace{-0.8in}
  \caption{The convex cost piecewise function associated with arc $(i,j)$.  }\label{fig:convex-duration}
\end{figure}

In the convex cost $K$-flow problem the cost functions are convex and piecewise linear, with integer piece sizes greater or equal to $1$.  Equivalently, for each arc $(i,j)\in A$ we have up to $K$ arcs between $i$ and $j$ with non-decreasing costs.  Each convex piecewise linear cost function $c_{ij}(x_{ij})$, has $k_{ij} \leq K$ integer breakpoints, $d_{ij}=b_{ij}^{(1)}> b_{ij}^{(2)} > \ldots >b_{ij}^{(k_{ij})}$,  and integer slopes (subgradients) $-w_{ij}^{(1)}, -w_{ij}^{(2)}, \ldots , -w_{ij}^{(k_{ij})}$.  The absolute values of the slopes of the linear pieces satisfy $w_{ij}^{(1)} < w_{ij}^{(2)} < \ldots < w_{ij}^{(k_{ij})}$, so the functions are convex monotone non-increasing.  An illustration of such function, for arc $(i,j)$, is given in Figure \ref{fig:convex-duration}.

As indicated in  Section \ref{sec:crashing}, the arc capacities in the crashing graph are determined by the right and left subgradients of the cost function $c_{ij}(x_{ij})$ of the arc at $x_{ij}$:
$c_{ij}^+(x_{ij})=c_{ij}(x_{ij})-c_{ij}(x_{ij}+1)$ and, $c_{ij}^-(x_{ij})=c_{ij}(x_{ij}-1)-c_{ij}(x_{ij})$.  These are equal to consecutive values of $w_{ij}^{(\ell)}$, $w_{ij}^{(\ell +1)}$, or if both reside on the same linear piece, then these are both equal to some $w_{ij}^{(\ell)}$.  The capacity lower and upper bounds in the crashing graph, are $(c_{ij}^+, c_{ij}^-)$.  For a convex cost function $c_{ij}(x_{ij})$, the capacity bounds are set as follows:

\[ (\ell _{ij}(\x), u_{ij}(\x))= \left\{ \begin{array}{ll}
     (0, 0)  & \mbox{ if     }\   x_{ij}= d_{ij} \ {\rm and }\  s_{ij}(\x) \geq 1\\
      (0, c_{ij}^-(x_{ij}))  & \mbox{ if     }\  \underline{d} _{ij} <  x_{ij}= d_{ij}\  {\rm and }\  s_{ij}(\x)=0\\
      (c_{ij}^+(x_{ij}), c_{ij}^-(x_{ij})) & \mbox{ if     }\ \underline{d} _{ij} <  x_{ij}< d_{ij}.
                \end{array}
\right. \]

In our case, for the TCT corresponding to the flow problem $\underline{d} _{ij} =-\infty$ for all activity arcs and therefore the two cases conditioned on $x_{ij}=\underline{d} _{ij}$ for the general TCT cannot occur for the flow problem.  

As for the bottleneck value, it depends on the length of the linear pieces, in the piecewise linear functions of the arcs on the cut:

\[ \delta = \min \left\{ \begin{array}{ll}
x_{ij} -b^{(p+1)}_{ij}& \mbox{ if     }\ ( i,j)\in (S,T),\ (i,j)\in A_{c}^0 \ b^{(p+1)}_{ij}< x_{ij}\leq  b^{(p)}_{ij}\\
b^{(p)}_{ij} -x_{ij} & \mbox{ if     }\ ( i,j)\in (T,S) {\mbox {\ and }} b^{(p+1)}_{ij}\leq x_{ij} < b^{(p)}_{ij}  {\mbox {\ and }} b^{(p)}_{ij} \leq d_{ij}\\
      s_{ij} & \mbox{ if     }\ ( i,j)\in (S,T),\ (i,j)\in A_{c}^+.
                \end{array}
\right. \]

\subsection{Example assignment problem}\label{sec:assignment}
\begin{figure}[h!]
		\hspace{-0.45in}
		\centering
\begin{subfigure}[t]{0.5\textwidth}
			\centering
			\includegraphics[height=13cm]{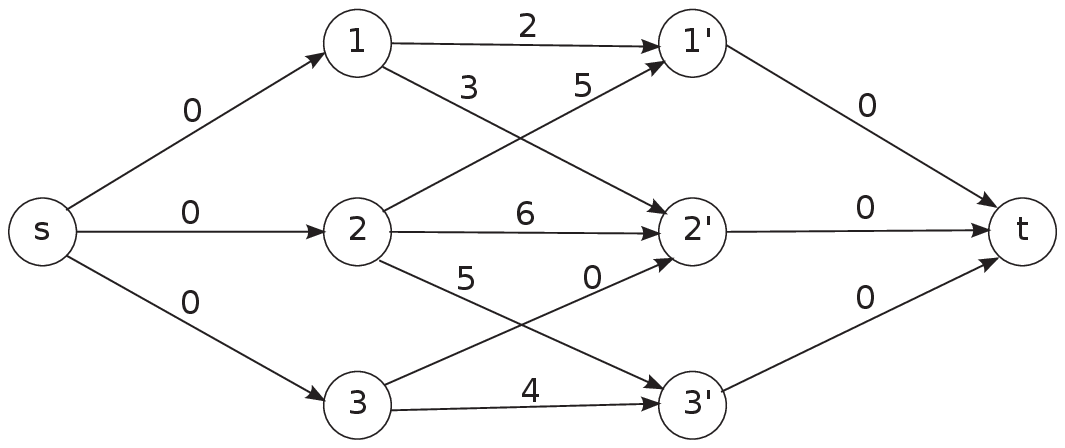}
\vspace{-2.98in}
			\caption{Assignment problem with arc durations/weights.}
		\end{subfigure}
		~ 
		\begin{subfigure}[t]{0.5\textwidth}
			\centering
			\includegraphics[height=13cm]{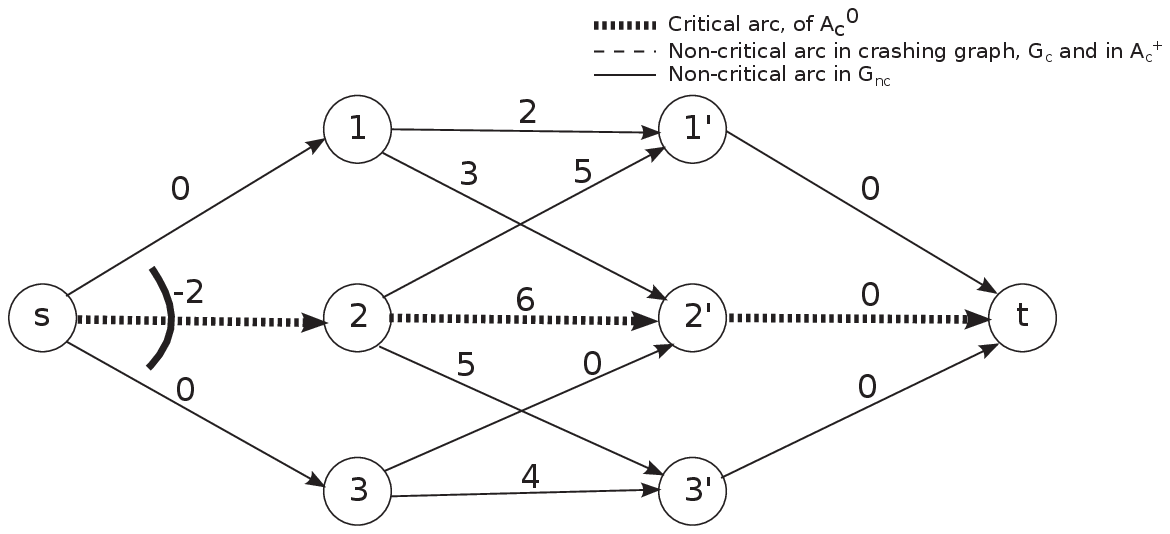}
\vspace{-2.98in}
			\caption{Crashing graph and durations' updates at iteration 1.}
		\end{subfigure}
\vspace{-2.607in}
\caption{First iteration} 
\label{fig:it0-1}
	\end{figure}

We present here the application of the algorithm for solving the assignment problem.  The problem is stated as maximizing the utility of assigning $n$ people to $n$ jobs.  In the $s,t$ bipartite network representing the problem, the arcs adjacent to $s$ and the arcs adjacent to $t$ have weight (cost) of $0$ and capacity of $1$.  All arcs $(i,j)$ in the bipartition have weight of $w_{ij}$ and capacity $\infty$.
The example presented has $n=3$ given with the arc weights shown in Figure \ref{fig:it0-1}(A).  The outcome of the first iteration is given in Figure \ref{fig:it0-1}(B).

\begin{figure}[!h]
		\hspace{-0.45in}
		\centering
\begin{subfigure}[t]{0.5\textwidth}
			\centering
			\includegraphics[height=12cm]{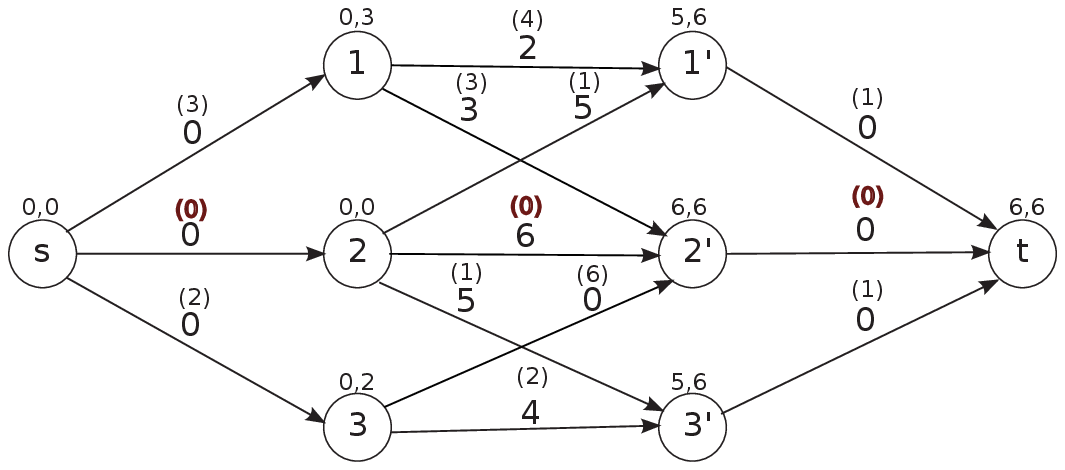}
\vspace{-2.9in}
			\caption{Applying CPM to evaluate $ET_j$, $LT_j$, and $(slack)$s.}
		\end{subfigure}
		~ 
		\begin{subfigure}[t]{0.5\textwidth}
			\centering
			\includegraphics[height=12cm]{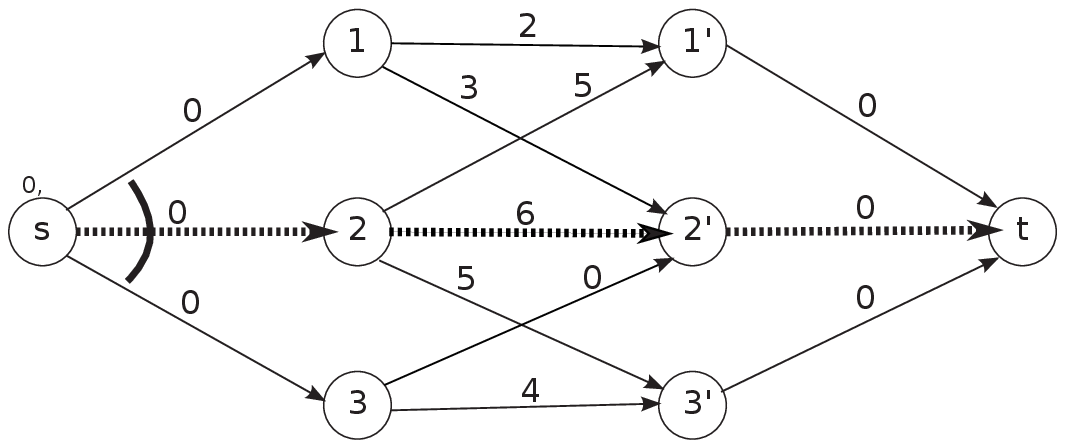}
\vspace{-2.9in}
			\caption{First cut in crashing graph with max-flow $=1$.}
		\end{subfigure}
\vspace{-2.526in}
\caption{CPM for iteration 1.}
\label{fig:CPM}
	\end{figure}

\begin{figure}[!h]
		\hspace{-0.45in}
		\centering
\begin{subfigure}[t]{0.5\textwidth}
			\centering
			\includegraphics[height=12cm]{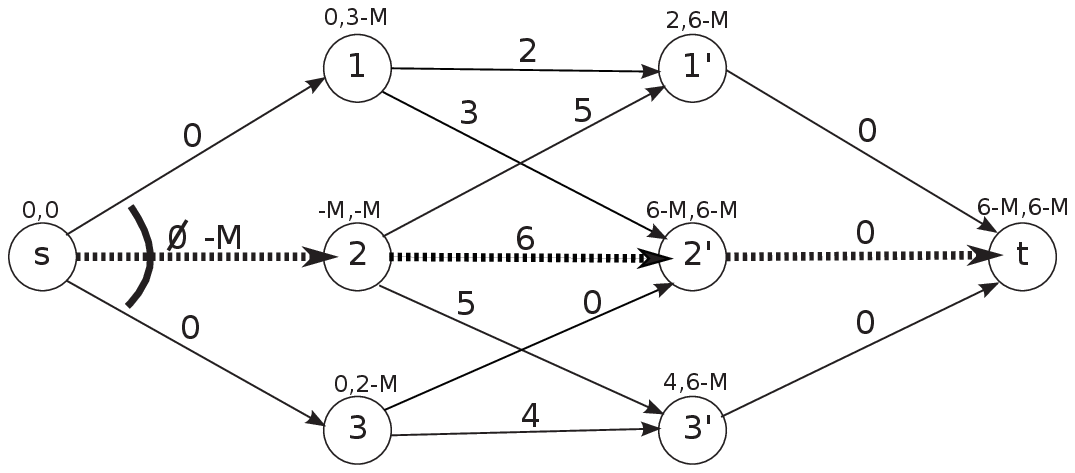}
\vspace{-2.98in}
			\caption{Applying CPM to update $ET_j$, $LT_j$.}
		\end{subfigure}
		~ 
		\begin{subfigure}[t]{0.5\textwidth}
			\centering
			\includegraphics[height=12cm]{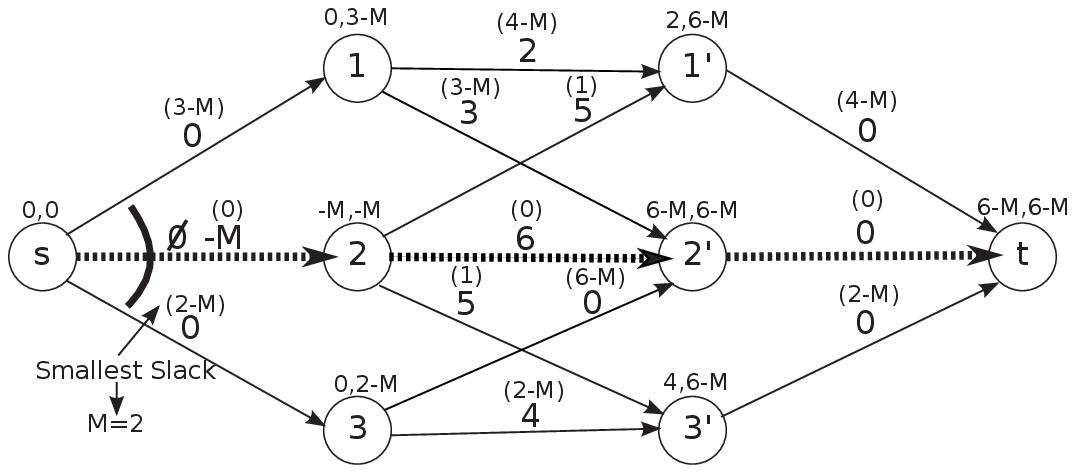}
\vspace{-2.98in}
			\caption{Determining smallest slack and duration reduction $M$.}
		\end{subfigure}
\vspace{-2.607in}
\caption{Applying {\sc find-critical}.}
\label{fig:find-critical}
	\end{figure}

The intermediate steps for iteration 1 are shown in Figure \ref{fig:CPM}: In (A) the crashing graph is determined by applying CPM.  It consists of the critical path $(s,2,2',t)$. The maximum flow in this crashing graph is of value $1$ and the cut is the arc $(s,2)$ of bottleneck value $M$.  Since this bottleneck value is infinite, the all-min-cuts stage for iteration 1 terminates with procedure {\sc find-critical} illustrated in Figure \ref{fig:find-critical}.

\newpage


\vspace{0.01in}
Figure \ref{fig:find-critical} demonstrates how to apply {\sc find-critical} and determine that $M=2$ and that nodes $3$ and $3'$ are consequently added as critical nodes.

\begin{figure}[!h]
		\hspace{-0.45in}
		\centering
\begin{subfigure}[t]{0.5\textwidth}
			\centering
			\includegraphics[height=12cm]{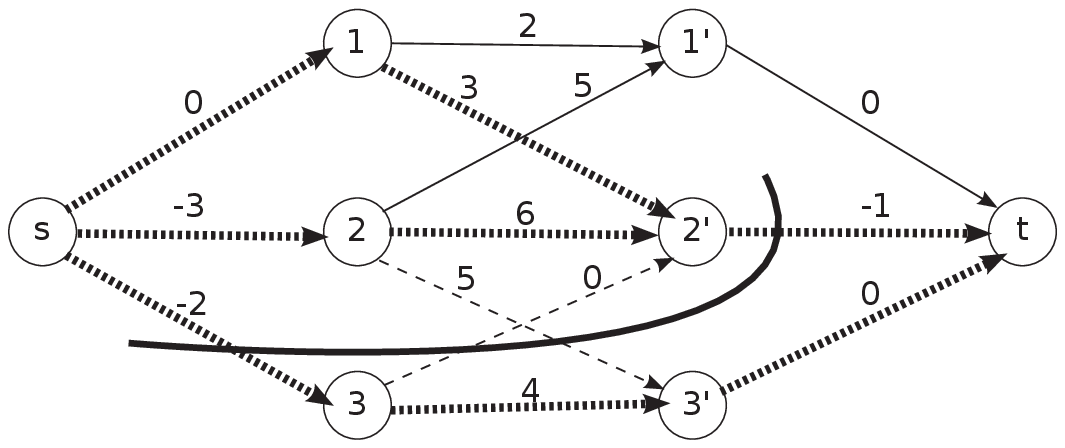}
\vspace{-2.98in}
			\caption{First cut in iteration 2, of bottleneck $1$.}
		\end{subfigure}
		~ 
		\begin{subfigure}[t]{0.5\textwidth}
			\centering
			\includegraphics[height=12cm]{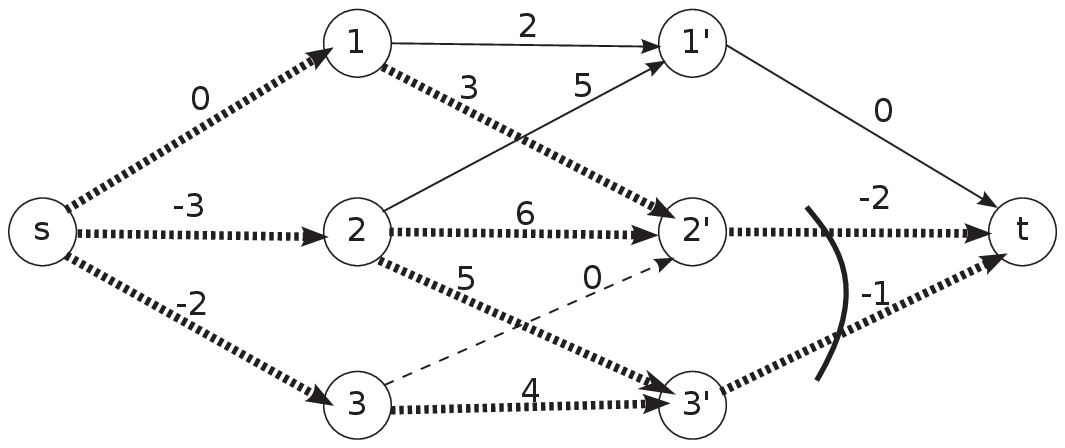}
\vspace{-2.98in}
			\caption{Second cut in iteration 2.}
		\end{subfigure}
\vspace{-2.607in}
\caption{The two cuts, of value $2$, in iteration 2.}
\label{fig:iter2}
	\end{figure}

In Figure \ref{fig:iter2} the sequence of two cuts in iteration 2 is shown.  Both these cuts are of capacity $2$.  For the first one, the bottleneck is attain for a cut-forward arc of $A_c^+$ of slack value $1$.  The second cut has bottleneck value of $M$, which is found to be equal to $1$.

\begin{figure}[!h]
		\centering
\begin{subfigure}[t]{0.5\textwidth}
			\centering
			\includegraphics[height=12cm]{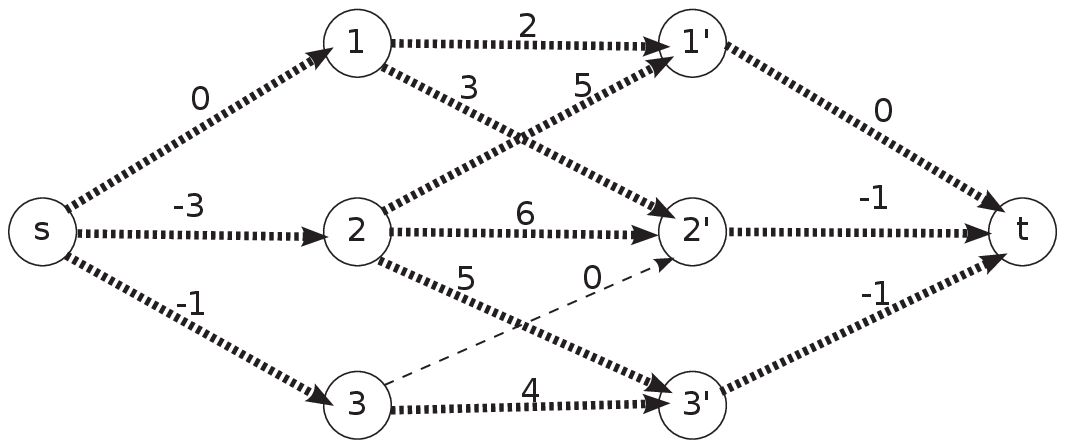}
\vspace{-2.98in}
			\caption{The final crashing graph with max flow $3$.}
		\end{subfigure}
		~ 
		\begin{subfigure}[t]{0.5\textwidth}
			\centering
			\includegraphics[height=12cm]{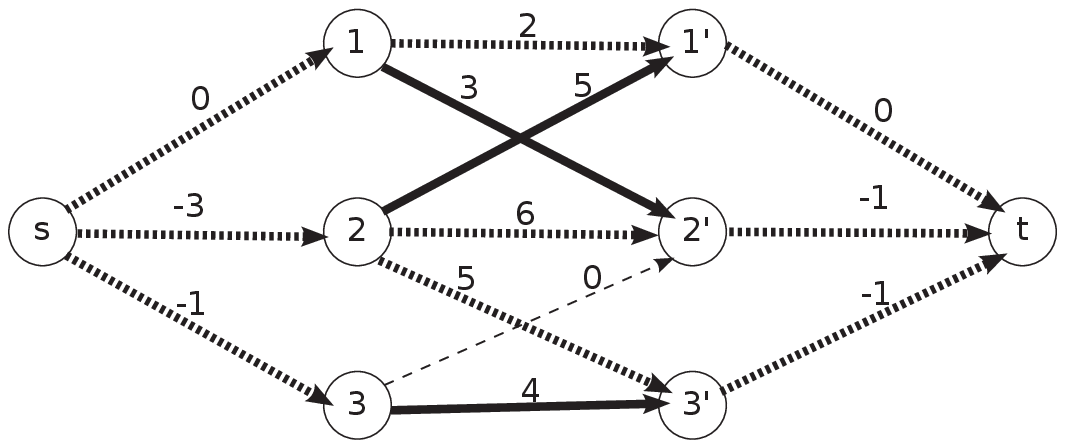}
\vspace{-2.98in}
			\caption{One of two optimal assignments corresponding to a max flow in the final crashing graph.}
		\end{subfigure}
\vspace{-2.607in}
\caption{The final crashing graph with max flow $3$.}
\label{fig:final}
	\end{figure}

The crashing graph at the beginning of iteration 3 is shown in Figure \ref{fig:final}.  The maximum flow in this crashing graph is $3$, which triggers the termination of the algorithm.  Any flow of value $3$  in this crashing graph corresponds to an optimal assignment.  In this example the solution is not unique.

\section{Warm start with node potentials, and accommodating cyclic graphs}\label{sec:warmstart}
\subsection{Warm start}
One can use an initialization to speed up the AMC algorithm.  For bipartite networks that include the assignment problem, transportation  problem, and weighted b-matching problem (the latter equivalent to assignment problem) we demonstrate here an initialization with a feasible solution that makes all nodes critical.  The use of this initialization eliminates the calls for procedure {\sc find-critical} and consequently no updates of the crashing graph are needed.

For a bipartite graph $(V_1\cup V_2,A)$ and arc weights $w_{ij}$, let $v_j =\max _{(i,j)\in A}w_{ij}$ for each $j\in V_2$; and $u_i =\max _{(i,j)\in A} \{ w_{ij}-v_j\}\leq 0$ for each $i\in V_1$.  We assign to each sink adjacent arc $(j,t)$ the ``duration" of $-v_j$, and for each source adjacent arc $(s,i)$ the ``duration" of $-u_i$.

With these durations, the $ES(i)$ values are equal to $-u_i$ and the $ES(j)$ values are equal to $\max \{-u_i+w_{ij}\} = v_j$.  The project finish time, and longest path from $s$ to $t$, is then $0$.   Computing the $LS()$ values of all nodes with the reverse dynamic programming, they all match the $ES()$  values and therefore all nodes are critical.  For each node of $V_2$ there is at least one critical incoming arc, and for each node of $V_1$ there is at least one critical outgoing arc.

Such an assignment of durations is a valid feasible solution, as follows from the next theorem:
\begin{thm}
For any $K$-MCF, assigning scalar cost values $\alpha _i$ to arcs adjacent to supply nodes (and source), $(s,i)$ and $\beta _j$ to arcs adjacent to demand nodes (and sink) $(j,t)$, results in an equivalent problem.
\end{thm}
{\bf Proof:} The assignment of the $\alpha _i$ and $\beta _j$ values adds to the objective the cost of $\sum _i \alpha _i s_i +\sum _i \beta _j d_j$ for $s_i$ $d_j$ the respective supplies and demands of nodes $i$ and $j$.  This quantity is a constant and therefore does not modify the solution.


\qed

Each path from $s$ to $t$ is now of length
$w_{ij}-v_j +|u_i| \leq 0$ (this is for longest.  for shortest we use $-w_{ij}$ and reduced cost of $-w_{ij}-|u_i|+v_j \leq 0$).  We let $|u_i|$ and $v_j$ be the node potentials.  This preserves the relative length of all the path from $s$ to $t$, which both have node potential of $0$.

For general K-flow on a graph that is not necessarily bipartite, we find the longest (shortest) paths from $s$ to all nodes, in particular to all demand nodes.  Let these distances be $v_j$ for each demand node and assign $-v_j$ cost to arc $(s,j)$.  This makes the project finish time equal to $0$.  Then find the longest (shortest) paths from $t$, along backward arcs, to all supply nodes and let these labels be $u_i$ for each supply node $i$, and assign $-u_i$ costs to all arcs $(s,i)$. This guarantees that all supply and demand nodes are critical.  But other nodes in the network may still remain non-critical.  Hence this warm start will speed up the algorithm for any graph, but for general graphs there are still calls to {\sc find-critical} and possible subsequent updates to the crashing graph.

\subsection{Finding shortest paths in non-DAG networks} \label{sec:node-potential}
Let $\pi _i$ be the shortest paths distances from $s$ to all nodes $i\in V$.  These are found with one application of Bellman-Ford algorithm with complexity $O(nm)$.  One then replaces all arc costs $c_{ij}$ by $c_{ij}+\pi _i -\pi _j$, the so-called ``reduced costs" that are by definition of shortest paths labels non-negative.  Since the reduced costs graph has non-negative arc costs, solving the single source shortest paths problem in that graph is possible with Dijkstra's algorithm, at complexity of $O(m+n\log n)$ \cite{FT}.  As discussed above, this is equivalent to working with slacks rather than the original durations.

\section{Conclusions}\label{sec:conclusions}
We present here a new algorithm,  the AMC-algorithm (All-Min-Cuts), for solving the TCT problem of minimum cost expediting that reduces the finish time of the project at maximum net benefit. For any such problem that has up to $K$ breakpoints in the TCT curve (or up to $K$ different minimum expediting costs in a repeated cuts algorithm) the complexity is $O((n+K)(m+n\log n))$. AMC-algorithm is notable as it solves also the minimum cost $K$-flow problem in complexity of  $O((n+K)(m+n\log n))$, presenting the first known significant speed-up to several classes of minimum cost flow problems.  For example, AMC-algorithm the first strongly polynomial algorithm for the minimum cost flow in unit capacity network that improves substantially on the general minimum cost flow algorithm.

Well known problems for which the AMC-algorithm runs in improved complexity compared to best known to date, include the uniform costs TCT problems, where for $n'$ activity arcs the complexity of the algorithm is $O((n+n')(m+n\log n))$, and the minimum cost flow in unit, vertex or arc, capacity network of complexities $O(n(m+n\log n))$ and $O(m(m+n\log n))$ respectively.  For the minimum cost $K$-flow and the convex minimum cost $K$-flow the complexity of the algorithm is $O((n+K)(m+n\log n))$.

The AMC-algorithm is an alternative algorithm for the assignment problem that matches the complexity of the best known algorithm to date, $O(n(m+n\log n))$, but with the use of a very different approach.  The approach of the AMC-algorithm relies on the finding of all minimum capacity cuts in the graph that have the same value in $O(m+n\log n)$ for a graph on $n$ nodes and $m$ arcs.  This procedure may prove powerful in other contexts.  In the context of min cost flow problems, the approach demonstrates that the total amount of flow in the graph can be exploited to improve the complexity of algorithms.  Also, if the cuts in the flow graphs have capacities that can assume only a bounded number of values, improved algorithms may result as well.  Since capacity scaling algorithms result in bounded amounts of flow on the arcs, and therefore in the network, our new methodology can lead to new efficient capacity scaling algorithms for general minimum cost flow problems.

As the AMC-algorithm's methodology departs dramatically from any methods currently in use, it affords new and fresh insights in the well studied area of minimum cost flow problems, with a potential for improvements in complexities.

\end{document}